\documentclass[a4paper,12pt]{article}
\usepackage[margin=2.5cm,nohead]{geometry}
\usepackage[utf8]{inputenc}
\usepackage{natbib}
\usepackage{color}
\usepackage{sgame}
\usepackage{pdflscape}
\usepackage{graphicx}
\usepackage{amssymb}
\usepackage{amsbsy}
\usepackage{amsmath,amsthm}
\usepackage{booktabs}
\usepackage{rotating}
\usepackage{colortbl}
\usepackage{caption}
\usepackage{setspace}
\usepackage{wasysym}
\usepackage{longtable}

\usepackage{subcaption}
\usepackage{float}
\usepackage{graphicx}
\usepackage{soul}
\usepackage{hyperref}
\usepackage{multibib}
\newcites{supp}{Supplementary References}

\usepackage[nobottomtitles]{titlesec}
\begin{document}
\sloppy
\title{Predicting Political Ideology from Digital Footprints}

\author{Michael Kitchener\thanks{%
SoDa Laboratories, Monash University} 
\hspace{5mm} Nandini Anantharama\thanks{%
SoDa Laboratories, Monash University}\\
\hspace{5mm} Simon Angus\thanks{ Department of Economics  and SoDa Laboratories, Monash University}
\hspace{5mm} Paul A.~Raschky\thanks{%
\textit{Corresponding author}; Department of Economics  and SoDa Laboratories, Monash University;  email: paul.raschky@monash.edu.}}
\date{May 2022}
\maketitle
 

\begin{abstract}
\medskip 
\noindent This paper proposes a new method to predict individual political ideology from digital footprints on one of the world's largest online discussion forum. We compiled a unique data set from the online discussion forum reddit that contains information on the political ideology of around 91,000 users as well as records of their comment frequency and the comments' text corpus in over 190,000 different subforums of interest. Applying a set of statistical learning approaches, we show that information about activity in non-political discussion forums alone, can very accurately predict a user's political ideology. Depending on the model, we are able to predict the economic dimension of ideology with an accuracy of up to 90.63\% and the social dimension with an accuracy of up to 83.09\%. In comparison, using the textual features from actual comments does not improve predictive accuracy. Our paper highlights the importance of revealed digital behaviour to complement stated preferences from digital communication when analysing human preferences and behaviour using online data.\\\medskip  



\noindent\emph{Keywords:} data mining, political ideology, digital footprint, Reddit\\
\medskip 
\end{abstract}
\pagebreak
\onehalfspacing

\section{Introduction}
\label{Introduction}

\noindent
Every activity online -- a post, a tweet, a comment, or a subscription to a discussion forum -- leaves a unique set of individual traces and in many instances, these traces are stored indefinitely and publicly accessible.  The infamous Facebook-Cambridge Analytica data scandal\footnote{Cambridge Analytica boasts of dirty tricks to swing elections. \textit{The Guardian}, 2018, \url{https://www.theguardian.com/uk-news/2018/mar/19/cambridge-analytica-execs-boast-dirty-tricks-honey-traps-elections} (accessed 21 October 2021).} has shown to the world, the potential that our seemingly innocuous online behaviour can be used to predict personal traits and link them to political preferences to ultimately  influence the outcome of elections.


While, there is a growing body of academic work devoted to predicting the personal traits of individuals (personality, gender, sexuality, etc.) on the basis of their digital footprints \cite[e.g.][]{Ross2009, Amichai-Hamburger2010,10.1145/2380718.2380722,kosinski2013private,Golbeck2011a, Gosling2011,wang2018deep}, directly predicting political ideology from digital footprints has received comparatively little attention. Existing studies have mainly relied relied on data explicitly political, \emph{stated preferences} in digital footprints, such as text comments or hashstags from twitter \cite[e.g.][]{Conover2011,Makazhanov2013, Boutet2012,Barbera2015} or likes on facebook \cite[e.g.][]{kosinski2013private}. 

We complement this literature by focusing on arguably non-political,  \emph{revealed preferences} in digital footprints and their ability to predict political ideology. This paper leverages an original data set to predict economic and `social' ideology, using textual and non-textual digital footprints, in a multi-class classification problem that allows for centrist self-identification. We have scraped the usernames and political ideologies of 91,000 users active on Reddit, one of the largest social news and discussion platforms globally. We have also recorded the frequency with which (most of) these users comment and post in a range of interest based discussion sub-forums (`subreddits') as well as the textual content of around 100 comments per user. With this data, we model the political ideology of these users as a function of the extent to which they engage with different subreddits and the textual content of their comments.  

Compared to existing studies, our dataset has two unique features. First, our data on political ideology not only distinguishes between and economic and social dimension of the left-right spectrum, but also allows for centrist self-identification (on both the economic and `social'\footnote{The term `social' for this ideological dimension is somewhat misleading, we provide a fuller explanation of what is meant here in Section \ref{Data}.} dimensions of ideology). Second, for each user we have a measure of revealed (interaction in non-political subforums) and stated (text corpus of actual comments in each subforum) digital preferences. These aspects of the data allow us to unpack the nexus between digital footprints and political ideology in a number of dimensions. We commence our analysis by testing if the accurate prediction of ideology from digital footprints is robust to the inclusion of self-identified centrists. In a similar vein, we also illustrate how the quality of predictions is affected by the complexity of the ideological variable we seek to model. Further, we decompose users' ideologies into both an economic component and `social' component, allowing us to investigate whether the economic component of political ideology is more readily predictable than the `social' component. Finally, we model ideology using 1) the frequency of users' interactions with subreddits, 2) the textual content of their texts and 3) the union of these sets of features. 
We split our into training (64\%), validation (16\%) and testing (20\%) sets and use a set of standard statistical learning approaches, ZeroR, Random Forest, AdaBoost, multinomial logit, logit, to predict user's political ideology based on their interactions with particular discussion forums and the text content of their comments (as well as a combination of both).

We find that information about the user's activity in non-political discussion forums alone, can very accurately predict political ideology. Depending on the model, we are able to predict the economic dimension of ideology with an accuracy of up to 90.63\% and the social dimension with and accuracy of up to 82.02\%. We also show that classification of a complex dependent variable representing nine different ideological classes can be achieved with a level of accuracy above baseline levels. In all cases, the frequency of interactions with different subreddits (analogous to digital footprints like search logs, purchase histories, etc.) is a stronger predictor than the textual content of comments and the union of both sets of features only marginally improves performance.

We note that modelling ideology as a function of digital footprints is, in part, provided for by the fact that an individual’s latent psychological makeup drives both their online behaviour and their political beliefs. Importantly, this provides for the observed association between online behaviour that is not of an explicitly political nature and ideological views (posting in Marxist or conservative discussion forums would constitute explicitly political online behaviour). It is the fact that digital behaviour of an entirely non-political nature can imply ideology that is cause for concern; for instance, as we show later,  frequently using the word `feel' in online comments is strongly associated with economically left wing views.  This is central to the privacy concerns we wish to illuminate.  

Our paper contributes to various literature branches within the broader area of ‘social data science’ focused on predicting private traits from digital data.

There is a growing body of work illustrating how online behaviour implicitly discloses private traits. Pioneers in linking traits to digital behaviour relied on volunteers granting access to their social media information and were thus constrained to relatively small sample sizes    \citep{Ross2009, Amichai-Hamburger2010, Golbeck2011a, Gosling2011}. More recently, researchers have begun using larger data sets with thousands of users or posts as the basis of analysis to make more robust findings. It has been shown that sexual orientation can be accurately predicted from pictures of an individual’s face with neural networks   \citep{wang2018deep} and that Facebook Likes can accurately predict Big Five personality traits, gender, race, drug use and other traits   \citep{kosinski2013private}. Even high level features of a Facebook profile (number of friends, number of statuses posted) have been shown to correlate with and predict personality   \citep{10.1145/2380718.2380722}. 

Indeed, it has been found that digital footprints can be used to train statistical models capable of predicting people’s personality traits to a higher degree of accuracy than their close friends and family   \citep{youyou2015computer}. Publicly available information from Twitter profiles   \citep{Golbeck2011, Quercia2011} and Facebook messages   \citep{Schwartz2013} has also been used to predict these traits using natural language processing techniques. Meta analyses claim that “the predictive power of digital footprints over personality traits is in line with the standard “correlational upper-limit” for behavior to predict personality”   \citep{Azucar2018} and that psychosocial traits more broadly can be accurately predicted by digital footprints   \citep{settanni2018predicting}. 

Further, it is not only digital traces from social media that can be effectively employed to predict private traits. The word choices of blog writers   \citep{Yarkoni2010},  the images `favorited' by Flickr users   \citep{Segalin2017} and mobile phone usage data   \citep{stachl2020predicting} have also be used to predict personality traits. 

Understanding the extent to which such prediction is possible informs the appropriate level of concern that should be paid to matters of digital privacy policy. Existing work has largely focused on psychological (e.g. Big Five personality traits) and demographic traits (e.g. gender). Pioneers in linking traits to digital behaviour relied on volunteers granting access to their social media information and were thus constrained to relatively small sample sizes  \citep{Ross2009, Amichai-Hamburger2010, Golbeck2011a, Gosling2011}. More recently, researchers have begun using larger data sets with thousands of users or posts as the basis of analysis to make more robust findings. It has been shown that sexual orientation can be accurately predicted from pictures of an individual’s face using neural networks   \citep{wang2018deep} and that Facebook Likes can accurately predict Big Five personality traits, gender, race, drug use and other traits   \citep{kosinski2013private}. Even high level features of a Facebook profile (number of friends, number of statuses posted) have been shown to correlate with and predict personality \citep{10.1145/2380718.2380722}. 

Indeed, it has been found that digital footprints can be used to train statistical models capable of predicting people’s personality traits to a higher degree of accuracy than their close friends and family   \citep{youyou2015computer}. Publicly available information from Twitter profiles   \citep{Golbeck2011, Quercia2011} and Facebook messages   \citep{Schwartz2013} has also been used to predict psychological traits using natural language processing techniques. Meta analyses claim that “the predictive power of digital footprints over personality traits is in line with the standard “correlational upper-limit” for behavior to predict personality”   \citep{Azucar2018} and that “digital traces from social media can be studied to assess and predict theoretically distant psycho-social characteristics with remarkable accuracy”   \citep{settanni2018predicting}. Further, it is not only digital traces from social media that can be effectively employed to predict private traits. The word choices of blog writers   \citep{Yarkoni2010},  the images `favorited' by Flickr users   \citep{Segalin2017} and mobile phone usage data   \citep{stachl2020predicting} have also been fruitfully applied to this end. These sorts of models can be effectively used to persuade people. Facebook advertisement with messages tailored towards predicted personality types, as indicated by Facebook Likes, increases engagement with advertised content   \citep{Matz2017}.
 

Political ideology has received relatively less focus. However, the accurate prediction of ideology from digital footprints could be misused to the detriment of society in several ways. If it is possible to accurately predict an individual’s ideology from their digital behaviour (even digital behaviour that is not of an explicitly political nature) then individuals with democratic sympathies living in authoritarian regimes may inadvertently reveal their non-conforming political views through ostensibly harmless online behaviour and be at risk of government retaliation. Additionally, the possibility of accurately estimating ideology may enable and encourage online political micro-targeting strategies aimed at voter suppression, which are considered to be in opposition with a functional liberal democracy by some \citep{anderson2018one}.  A number of studies have shown that natural language processing (NLP) techniques can be used to accurately classify the political leanings of Twitter users   \citep{colleoni2014echo, Conover2011, Makazhanov2013, Boutet2012} and their political engagement or ideological extremity   \citep{Preotiuc-Pietro2017}. Hashtags and network analyses of mention/retweet networks can also be leveraged to effectively classify the ideologies of Twitter users   \citep{Conover2011, Barbera2015, Boutet2012}. Kosinski et al.'s seminal work   \citep{kosinski2013private} pays some attention to political traits. Kosinski et al. map the Likes of 9,572 American Facebook users with a self-disclosed party affiliation of  either Democrat or Republican using a logistic regression model, illustrating that the model’s predictions on unseen data achieve an ROC-AUC (receiver operating characteristic, area under curve) of 0.85. Our approach to modelling ideology as a function of digital footprints is broadly based on the methods used in this paper. 

This body of work represents a great leap in our understanding of how online behaviour can be leveraged to predict ideology. However, some important questions remain unanswered. Firstly, the vast majority of these studies focus on predicting a binary variable representing ideology (for a systematic literature review see     \citep{Hinds2018}). In reality, ideology is not binary and many people identify as centrists so we want to know if the accurate prediction of ideology is robust to the inclusion of centrists. Some research has been done into multi-class prediction of ideology but these studies rely on textual   \citep{Boutet2012, Makazhanov2013, Preotiuc-Pietro2017} and network data   \citep{Boutet2012, Barbera2015}  which does not generalise to ubiquitous footprints like web search logs, and sometimes utilise fallible ground truths   \citep{Makazhanov2013, Preotiuc-Pietro2017}. Secondly, a one-dimensional conception of ideology is typically insufficient to capture people's political beliefs. Many people consider themselves socially progressive but fiscally conservative and cannot easily be placed into a simple left/right spectrum. Consequently, we want to know whether some dimensions of ideology are more readily predictable than others. We have not identified any studies that attempt to answer this question. Finally, it is unclear if textual or non-textual digital footprints (Facebook Likes, search history logs, etc.) are better predictors of ideology. Again, to our knowledge no study has directly compared the predictive power of textual and non-textual footprints that mimic a broader range of footprints.

Our research also illustrates the usefulness of Reddit data in the study of political ideology. Data from Reddit has been used to predict personality and gender   \citep{Gjurkovic2020}, and even detect mental illnesses like anxiety, depression and schizophrenia   \citep{Kim2020}. More broadly, there is growing literature on using digital footprints as revealed measures of preferences and human behaviour more broadly. For example, \cite{ackermann2017} and \cite{ackermann2020} use data from over a trillion observations of end-user internet connections and online/offline activity, to predict sleep and wake hours as well as electricity consumption. \cite{ NBERw27827} used publicly available data from over 2 billion Facebook users to create new measures of revealed cultural values and preferences and calculate cultural differences between countries. \cite{desmet21} employed data on Facebook interests to analyse the differences in preferences between genders across countries. \cite{NBERw29598} collected real-time data from public github profiles to estimate the effect of the COVID-19 pandemic and associated lock downs on labor activity. However, a systematic review of the literature on linking digital footprints to political orientation indicates that no one has previously used Reddit data to this end \citep{Hinds2018}. By showing Reddit to be a reliable source of rich social data we hope to spur further developments in this field. 



Our results also speak to the literature on the psychological foundations of ideology and of digital behaviour. There is a demonstrated association between psychological traits and certain interests   \citep{Wolfradt2001, McCrae1997}. Interests likely impact language use (if you like movies you probably use words like `actor' and `film' a lot) and people with certain interests are more likely to participate in online forums pertaining to these interests. 

Psychological traits also influence ideology. There is a substantive amount of literature in psychology that has focused on modeling ideology as a function of scores on particular psychometric scales. In one study, the Big Five model of personality was used to predict party preferences in Italy, Spain, Germany, Greece and Poland   \citep{Vecchione2011}. It was found that Big Five traits were more powerful predictors than demographic characteristics (age, gender, income, education) and that openness to experience predicts left wing views whilst conscientiousness predicts voting for conservative parties. They note that the linkage between personality and political views is consistent with theoretical expectations: conscientiousness comprises a preference for order and adherence with rules whereas openness to new ideas correlates against preferences for rigid social hierarchy   \citep{Vecchione2011}. It has been shown that including scores from the HEXACO model of personality as additional regressors alongside Big Five traits increases the proportion of variance explained when modeling ideology   \citep{Chirumbolo2010}. Another study illustrates that scores on Hartmann’s Boundaries Questionnaire can predict conservatism   \citep{Hiel2004}. Some results are consistent with theoretical expectations; conservatism is correlated with a preference for rigid and well defined social structures, whilst other results are harder to explain; conservatism is also positively correlated with a preference for blurred edges and soft lines in art. 

More broadly, a meta analysis of ``conservatism as a function of social cognition" has found that death anxiety, system instability, dogmatism, openness to experience, uncertainty tolerance, need for order, structure and closure, integrative complexity, fear of threat and loss and self-esteem all have predictive power of conservatism   \citep{Jost2003}. Consider how an individual’s interests and activities are likely influenced by these traits. Someone with substantial death anxiety may have a strong interest in existential philosophy. Someone with low openness to experience may prefer older, familiar television shows to more recent ones. Consequently, psychological traits influence both ideology and our interests, which in turn influences our online behaviour, providing for an observable mapping of online behaviour to ideology which can be approximated by statistical learning models.

The remainder of the paper is structured as follows: section \ref{Data} of this paper provides information on our data; section \ref{Methodology} details the modelling strategies; section \ref{Results} presents the results; section \ref{Discussion} concludes . 

\section{Data}
\label{Data}

To fill the gaps in the literature discussed in Section \ref{Introduction}, we require digital footprints and multi-dimensional ideological labels from a large number of internet users. Finding such data is not a trivial task. The MyPersonality data set that contains information on a large number of Facebook users' Facebook Likes is no longer available (nor does it contain sufficiently sophisticated ideological labels) and obtaining new data from Facebook is not realistic due to Facebook's increased protection of private data in response to privacy scandals. Twitter, whilst a viable source of data for such inquiry, is limited by the fact that there is no real way to effectively and easily collect information on the ideologies of large amounts of users. Further, using purely textual data (like Tweet data) will only allow us to predict ideology as a function of explicit textual communication. We would, however, also like to get an indication of how more subtle digital footprints that do not involve explicit communication can be used to predict ideology.

\begin{figure}[!htbp]
\centering
  \fbox{\includegraphics[width=10cm,height=4.7cm]{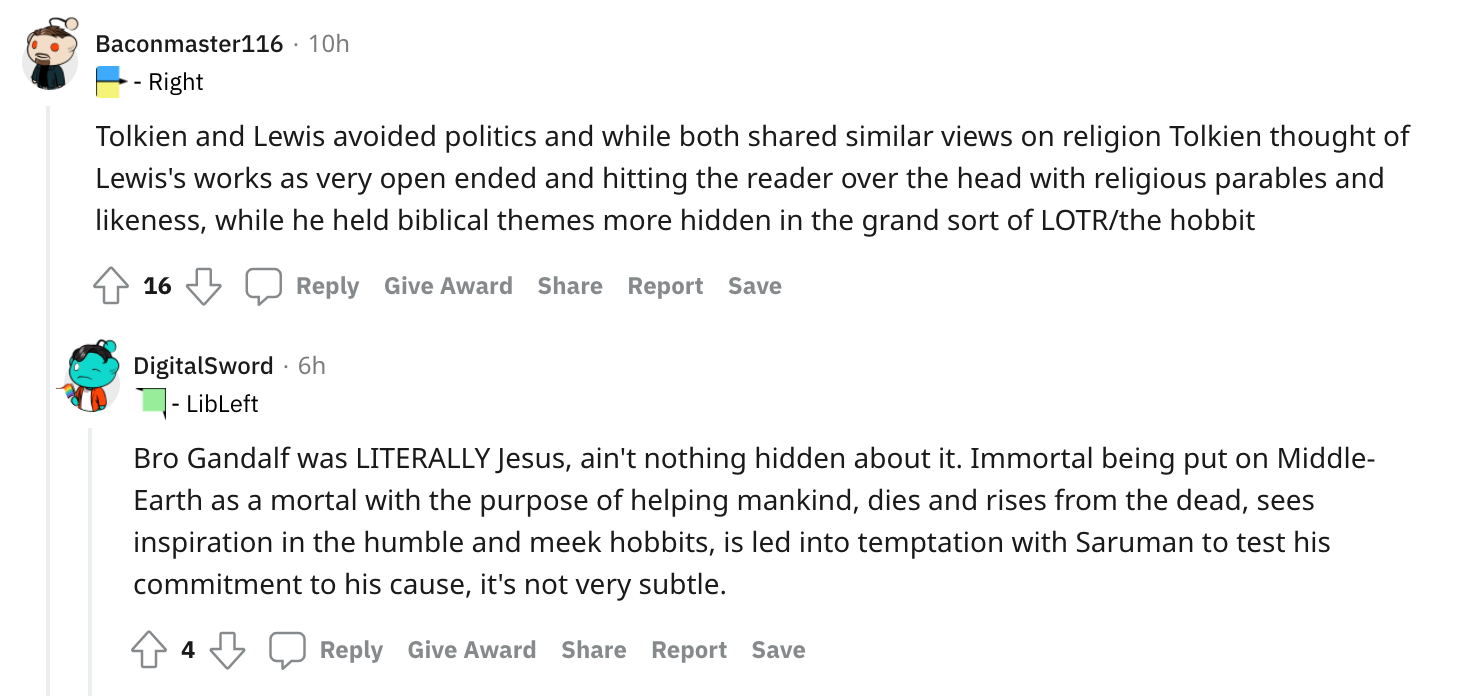}}%
  \caption{A comment from a user who has flaired themselves as having economically right wing views and a response from a user who has flaired themselves as having a `libertarian left' ideology \label{fig.pcm-comment-fig}}
\end{figure}

We opted to scrape an original data set from the  \href{https://www.reddit.com/r/PoliticalCompassMemes/}{r/PoliticalCompassMemes} subreddit, a Reddit forum in which users `humorously' critique ideologies in opposition to their own. In this subreddit, users ‘flair’ their posts with with their results from the popular \href{https://www.politicalcompass.org/test}{Political Compass Test}. We created a Python script using the Python Reddit API Wrapper (PRAW) library to scrape the r/PoliticalCompassMemes subreddit and collect the usernames, ideologies and post/comment histories of 91,000 users who had commented on posts in the subreddit. We also created a script using the PMAW Pushshift   \citep{Baumgartner2020} API wrapper to collect a maximum of 100 comments from each user\footnote{This data is available upon request.}. Figure \ref{fig.pcm-comment-fig} shows an example of how users' comments are flaired with their political ideology. 

\begin{figure}[!htbp]
\centering
  \fbox{\includegraphics[width=10cm,height=9.7cm]{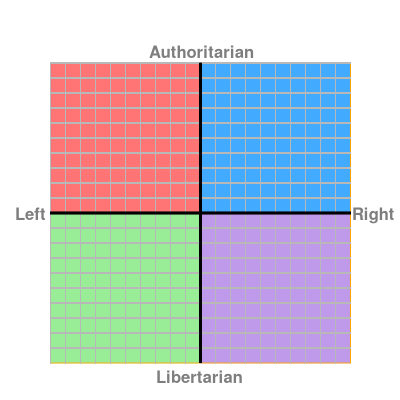}}%
  \caption{The political compass. Flaired users take a test specifying the extent to which they agree with politically charged statements (e.g. ``If economic globalisation is inevitable, it should primarily serve humanity rather than the interests of trans-national corporations") and are assigned to one of four quadrants: libertarian left (social democrat), libertarian right (libertarian), authoritarian left (communist) or authoritarian right (traditional conservative) depending on their answers \label{fig.policompass-fig}}
\end{figure}

Figure \ref{fig.policompass-fig} shows the political compass; users are assigned to one of the four quadrants in the two-dimensional ideological space. 

The data collection process was organised into three stages: First, we created a list of usernames and ideology signaling flairs associated with each username. Second, we compiled for each user their  going their post and comment history and recorded a tally of the amount of times they have posted and commented in each subreddit they interact with. Third, for each user, we collected the textual content of the 100 most recent comments they have made in subreddits excluding r/PoliticalCompassMemes. 

Kosinski et al. note that findings from studies using Facebook Likes likely generalise to other types of digital footprints   \citep{kosinski2013private}. The same is true of Reddit interactions. Reddit users are able to interact with millions of subreddits, which are interest or topic based discussion forums. As such, interaction with a subreddit can be seen as an expression of interest in the relevant topic (if a user interacts with the r/gaming subreddit, they are `showing' an interest in video games). Consequently, the frequency with which someone interacts with different subreddits is broadly analogous to other (likely richer) digital footprints such as web search logs (searching a topic also indicates interest in it)  so our results should generalise to the kinds of digital footprints that entities which threaten digital privacy have access to. 

\subsection{Creating a list of usernames and ideology signaling flairs}
\label{Data-creating-user-flair}

We created a list of usernames and associated ideology signaling flairs by running a Python script that cycled through the top 1,000 most popular posts of all time in the r/PoliticalCompassMemes subreddit. For each post, we looped through all the available comments. If the author of the comment was not already in a list of users whose ideology we had recorded and their comment was flaired with an ideology, we added their username and their flair as a row to the data set. The username was also added to the list of users whose ideology we had recorded to avoid doubling up. This resulted in a data set of 91,000 username and flair combinations. An illustration of this data can be found in Table \ref{table.user-flair-example}.

\begin{table}[!htbp]
\centering
\caption{Example user-flair data \label{table.user-flair-example}}
    \begin{tabular}{||c c ||} 
    \hline
    \textbf{username} & \textbf{ideology} \\ 
    \hline
    user1 &  Libertarian-Left \\
    \hline 
    user2 & Authoritarian-Right \\
    \hline
    user3 &  Libertarian-Right \\
    \hline
    \ldots & \ldots \\
    \hline
    \end{tabular}
\end{table}

\begin{figure}[!htbp]
\centering
  \fbox{\includegraphics[width=14cm,height=8.75cm]{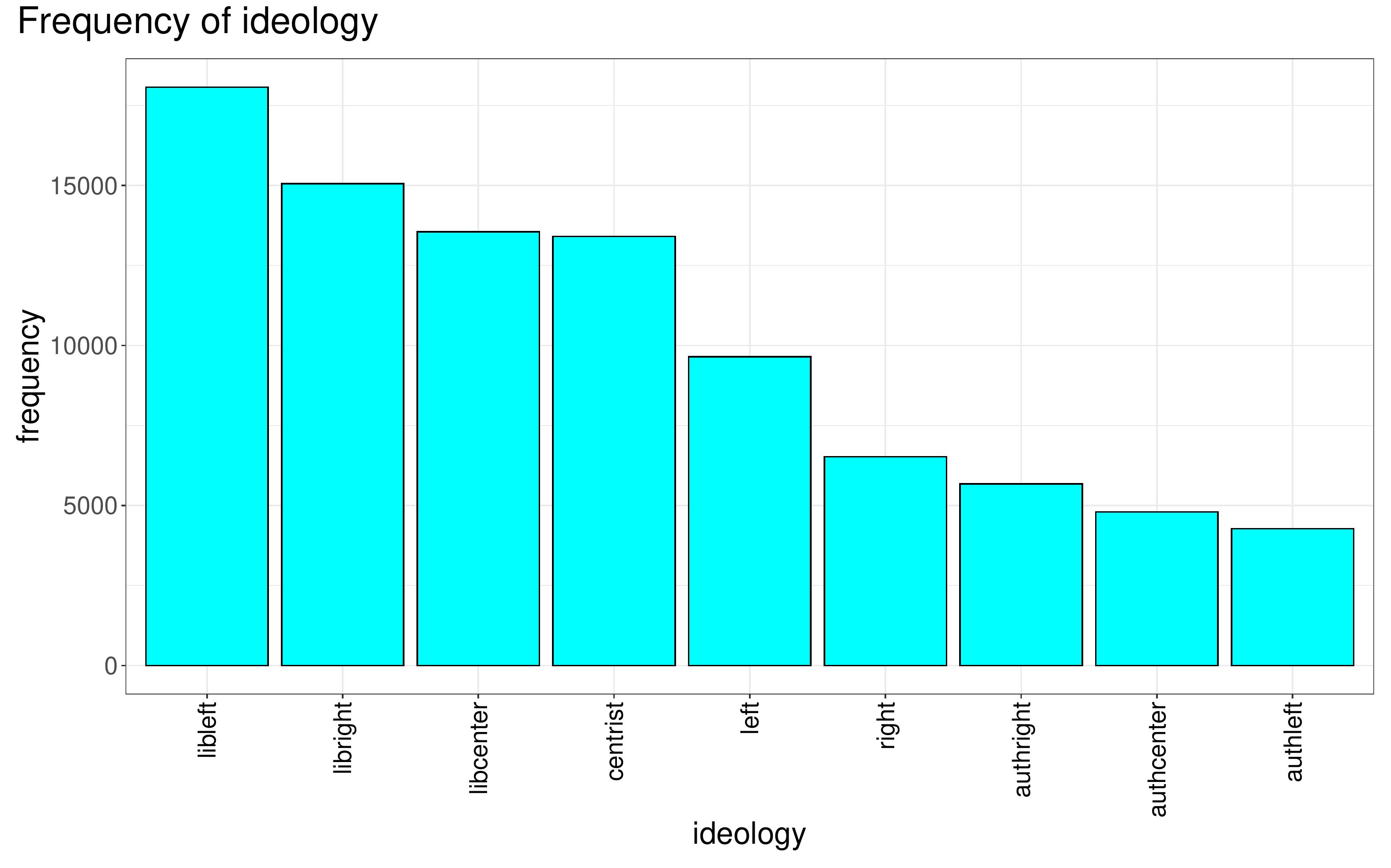}}%
  \caption{Ideology frequency in sample ($n = 91,000$), note: we were unable to gather footprints for some users so the size of the data set that is used in model training varies depending on the particular model \label{fig.response-freq}}
\end{figure}

There are a total of nine classes (different flairs that correspond to the same ideology were grouped together as detailed in Appendix \ref{Appendix-Recoding}). The \textbf{libleft} flair corresponds to social democrats, i.e. supporters of Bernie Sanders,  \textbf{libright} to libertarians, i.e. supporters of Ron Paul, \textbf{libcenter} to those who identify as libertarians with respect to the role of the state in social matters but have centrist economic views, \textbf{centrist} to  those who identify as having centrist views on the economy and the role of the state in social matters, \textbf{left} to those who identify as having left wing economic views but centrist views on the role of the state in social matters, \textbf{right} to those who identify as having right wing economic views but centrist views on the role of the state in social matters, \textbf{authright} to  traditional conservatives, i.e. supporters of Ted Cruz, \textbf{authleft} to communists, i.e. supporters of Xi Jinping and \textbf{authcenter} to those who identify as having authoritarian views with respect to the role of the state in social matters but centrist (or perhaps mixed) economic views\footnote{This ideology is treated as a pro-fascism ideology in the forum.}. The frequency and sample proportion of each ideological class is described in Table \ref{table.response-freq-prop} and illustrated in Figure \ref{fig.response-freq}.

\begin{table}[!htbp]
\centering
\caption{Ideology frequency and proportion in sample (n = 91,000) \label{table.response-freq-prop}}
\scalebox{0.8}{\begin{tabular}[t]{||r r r r r r r r r r||}
\hline
 & libleft & libright & libcenter & centrist & left & right & authright & authcenter & authleft\\
\hline
Frequency & 18,070 & 15,054 & 13,548 & 13,408 & 9,646 & 6,526 & 5,672 & 4,801 & 4,275\\
\hline
Proportion & 0.2 & 0.17 & 0.15 & 0.15 & 0.11 & 0.07 & 0.06 & 0.05 & 0.05\\
\hline
\end{tabular}}
\end{table}

The Political Compass Test only allocates users to one of four quadrants (authleft, authright, libleft, libright) by providing quantitative scores of how right or left they are economically and how libertarian or authoritarian they are `socially'. Presumably, users who identify as, say, `left' have taken the test and found that though they score highly `left' in the economic dimension, they are only marginally above or below the origin (midpoint) of the `social' axis. Likewise, centrist users may consider themselves insufficiently far from the origin in either dimension to warrant any other classification. Still, it is fundamentally up to the discretion of the user as to whether to identify in accordance with their test outcome or as a  centrist on one or both ideological dimensions. This choice may be based on prior beliefs regarding their political views. Consequently, the labels here must fundamentally be understood as self-reported ideologies rather than as results from the test even though these self-reported ideologies are presumably strongly influenced and calibrated by test results.

In what follows we refer to users' position on the left/right axis as economic ideology (position here is defined by responses to questions regarding the economy) and position on the authoritarian/libertarian axis as social ideology. It may not be immediately clear how the terms authoritarian/libertarian relate to `social' political ideology but inspection of the relevant questions on the Political Compass Test reveals that these questions elicit preferences for government's coercive involvement in social facets of life, i.e. whether the government should enforce a ban on abortion. As such the term `social' ideology provides a useful shorthand that will be employed in the remainder of this paper (with the omission of single quotation marks). 

There is no guarantee that each username in our data set corresponds to a unique person; one individual could have two or more different Reddit accounts. However, this is unlikely to be a prevalent behaviour and should have a negligible effect on the validity of our data. It is also possible that an account’s flair may not truly represent the ideology of the person behind the account. This could be due to users mistakenly assuming that they subscribe to a particular ideology without having taken the test to calibrate their self-assessment. It is also possible that some users create accounts flaired with ideologies other than their own in order to post stereotypes associated with adherents of that ideology in an effort to satirise the beliefs of their ideological opponents. We do not consider this a substantive issue in our data.

\subsection{Collecting the subreddit interaction records of flaired users}
\label{Data-collect-digital-footprint}

With the user-flair data collected, the next step was to obtain comprehensive digital footprints for all of the users whose ideology was known. For each user, we looped through their 1,000 most recent posts and 1,000 most recent comments. Each post or comment was stored as a row in a data set with the following columns: \textbf{user:} the username of the user whose posts/comments we are recording; \textbf{interaction:} whether the interaction being recorded is a comment or a post; \textbf{title:} the title of the post if the interaction is a post, otherwise blank; \textbf{score:} the `score' of the post/comment; other users can `upvote' a post or comment and increase its score by 1 or `downvote' the post or comment and decrease its score by 1; \textbf{time:} the time the post/comment was created; \textbf{subreddit:} the subreddit the post/comment was made in.

A hypothetical illustration of this data can be found in Table \ref{table.user-history-example}.

\begin{table}[!htbp]
\centering
\caption{Example user-history data \label{table.user-history-example}}
\scalebox{1}{\begin{tabular}{||c c c c c c||} 
\hline
\textbf{username} & \textbf{interaction} & \textbf{title} &  \textbf{score} & \textbf{time} & \textbf{subreddit} \\ 
\hline
user1 &  post & GTA V \dots &  43 & 10:32 2/1/21 & r/gaming  \\
\hline 
user1 &  comment &  & 12 & 16:12 8/12/20 & r/classicalmusic \\
\hline 
user2 &  post & Today's \dots &  -6 & 5:36 4/3/21 & r/boxing \\
\hline 
user2 &  post & The \dots &  3 & 4:24 4/3/21 & r/seinfeld \\
\hline 
user2 &  comment &  &  0 & 5:09 3/3/21 & r/crypto \\
\hline 
\ldots & \ldots & \ldots      & \ldots  & \ldots  & \ldots \\
\hline
\end{tabular}}
\end{table}

This can be transformed into a user-interaction matrix where each row represents a unique user and each column refers to a particular subreddit. The value in any particular cell can represent a number of things: the number of times the user has posted/commented in that subreddit, whether the user has posted/commented in that subreddit or the average score of the user’s posts/comments in that subreddit. Table \ref{table.user-interaction-example} illustrates what a user-interaction matrix may look like.

\begin{table}[!htbp]
\centering
\caption{Example user-interaction matrix \label{table.user-interaction-example}}
\begin{tabular}{||c c c c c||} 
\hline
\textbf{username} & \textbf{r/gaming} & \textbf{r/classicalmusic} & \textbf{r/boxing} & \textbf{r/seinfeld} \\ 
\hline
user1 &  3 & 12 & 0 & 0 \\
\hline 
user2 &  0 & 4 & 1 & 6 \\
\hline 
user3 &  0 & 0 & 0 & 0 \\
\hline 
user4 &  1 & 0 & 0 & 14 \\
\hline 
user5 &  0 & 43 & 0 & 0 \\
\hline 
\ldots & \ldots & \ldots   & \ldots   & \ldots  \\
\hline
\end{tabular}
\end{table}

The user-interaction matrix is a set of predictors for each user and can be merged with the user-flair data. This allows us to model ideology as a function of digital interactions. As described in Section \ref{Methodology-user-interaction}, to examine the extent to which ideology may inadvertently be revealed, we removed columns that represent interactions with explicitly political subreddits when developing predictive models. Since there are a huge number of (debatably) political subreddits we are unable to remove all of them, however, by removing the many of the most popular political subreddits we hope to minimise the influence of explicitly political subreddits on our models. 

The user-history data set is very large (63,709,041 rows) so transforming it into a user-interaction matrix is not a trivial computational task. The user-history data is pivoted chunk by chunk. The final user-interaction matrix comprises the union of all chunks. We transformed the user-history data set into a user-interaction matrix where each cell represents the amount of times the user has posted or commented in the relevant subreddit. 

It should be noted that the limit of 1,000 posts/comments is not arbitrary. PRAW limits requests to 1,000 objects at a given time; i.e. if we are looping through a particular subreddit’s top posts we can at most request 1,000 post objects from PRAW. It was on this basis that we elected to gather usernames and flairs through the top posts since these are likely to have many comments and hence more user/flair combinations to record. We also chose to record each user's 1,000 most recent posts and comments as these are presumably most reflective of the user's most current attitudes and interests. There may not be exactly 1,000 records returned from any request since deleted posts may be returned and contribute to the request limit despite being of no use to us. 

If our user-history script encountered an error whilst collecting records of subreddit interactions for a given user (this could happen occasionally, often due to issues with Reddit's servers) it disregarded all data collected for that user and skipped to the next user to avoid censoring records of interactions for specific users.

\subsection{Collecting the comments of flaired users}
\label{Data-collect-comments}

We looped through each user in our list and recorded their 100 most recent comments from any subreddit excluding r/PoliticalCompassMemes. The process of extracting features to use in our models from this raw textual data is described in Section \ref{Methodology-text}.

In the user-interaction matrix, we removed interactions with popular, explicitly political subreddits from our set of features. We do not remove explicitly political words (i.e. words that may be associated with partisanship) as there is no way to a priori determine how the use of politically charged words discloses ideology. For instance, a progressive may use the words ‘Trump’ and ‘gun’ a lot as they criticise the Trump administration and lack of stringent gun control in the United States. That is to say, the use of these politically charged words does not explicitly disclose an ideological viewpoint (as we assume participation in explicitly ideological subreddits to). Consequently, we do not remove any words from our corpus. 

We could have avoided scraping comments from our list of politically explicit subreddits but there is no reason to think that the language used in these subreddits constitutes an explicit disclosure of ideology. Knowing the textual content of a comment in an explicitly political subreddit without knowing what the subreddit actually is, is insufficient information to infer ideology. Consider the following analogy, if someone tells you that they spend last Thursday night at their local Labor Party branch meeting then they are effectively disclosing their political beliefs to you. If they tell you they spend last Thursday night talking about Federal Parliament and trade unions (whilst at a Labor Party branch meeting, though they do not tell you this) they have not definitively revealed their political beliefs.

The scraper occasionally encountered errors whilst collecting comment data. When this happened, we used an error catch to drop all data for the user whose data we were collecting when the error occurred (to avoid censoring the textual content of comments for specific users) and skipped to scraping the comments of the next user in our list.


\subsection{Caveats and Advantages of Online Discussion Forum Data}
\label{Data-advantages}

There are some caveats common to all research with digital data that should be mentioned. As noted, our subreddit interaction records for a given user are limited by Reddit’s API. We also only recorded the textual content of a maximum of 100 comments per user due to computational restrictions. Such issues are endemic to data scraped from the web where API limits often restrict how much data can feasibly be extracted for each user. 

Further, our sample is not representative of any particular national population. Users may come from any country (although most appear to come from English speaking countries). The distribution of key demographic variables like age, education, income and race in the population of Reddit users is likely far less varied than it is in any country level population. Further, users who comment and flair themselves in the r/PoliticalCompassMemes subreddit likely have different traits to the broader population of Reddit users. Clearly, there are major sample selection issues here. We are not attempting causal inference but there is a risk that the average r/PoliticalCompassMemes commentor has a greater interest in politics than the average Reddit user. This may manifest in greater engagement with activities (and hence subreddits) which are strongly associated with particular ideologies as ideology is a larger part of personal identity for these people. As such, our model may over estimate the predictive accuracy of digital footprints over political views and our findings/models may not generalise to predicting the ideologies of less politically engaged users whose interests (and digital interactions) are less tied to their political views. However, these selection effects may also deflate the observed associations between predictors and response, causing our models to understate the predictive power of digital footprints. The sample selection effects constrain the range of the distributions of our predictors (people who interact with the r/PoliticalCompassMemes subreddit likely have a narrower range of interests and hence subreddit use and language variation than the broader Reddit population) and accordingly reduce the variance in our predictors that can correlate with our response variable. We note that this sample selection issue is common in all studies focused on predicting private traits through digital records, many of which feature selection bias by requiring users to voluntarily disclose relevant traits   \citep{kosinski2013private, 10.1145/2380718.2380722, colleoni2014echo, youyou2015computer, Golbeck2011, Quercia2011, Preotiuc-Pietro2017}.

Another potential issue comes from how we collected users' ideologies as signalled by their flairs at time $t_1$ but scraped their most recent comments at time $t_2$ where $t_2 > t_1$. If a user's ideology has changed between $t_1$ and $t_2$ then their digital footprints may not reflect the digital behaviour associated with their flaired ideology, undermining predictive models. We do not consider this to be a substantive risk  as the subreddit has only been popular for a few years so there is typically a relatively short interval between the creation of flaired comments and the collection of digital footprints.

The previous section elaborates on the limitations of our data that are, by and large, endemic to all research in this field. However, our data also promises several advantages over that used in previous studies. We will describe the advantages of our data with reference to the MyPersonality data set used by Kosinski et al.   \citep{kosinski2013private} since this work provides the methodological basis of our paper. The MyPersonality data set features records of the Facebook Pages ‘Liked’ by over 58,000 Facebook Users who agreed to take an online personality survey and have data collected from their Facebook profile. The data set also contains the US party affiliation of 9,572 Facebook Users who declared this information on their profile. 

As noted, we want our data to proxy a broader range of digital footprints so that we can understand just how salient digital privacy concerns are. Facebook Likes and Reddit interactions data are both reasonable proxies of  web search logs, purchase records, and other digital footprints. However, we propose that Reddit data better emulates these footprints for several reasons.

Firstly, due to the anonymity of Reddit relative to Facebook, users may be more willing to implicitly disclose private interests via digital behaviour (users may comment in certain pornographic subreddits but few people would willingly disclose this information via their Facebook Likes). These private interests may correlate with ideology. The fact that Reddit data captures this kind of information makes it a better proxy: consider how people typically use search engines under the presumption that their search queries are private. Essentially, Reddit data expresses the same range of interests as other digital footprints whereas Facebook data only expresses a restricted range of interests. 

Further, the MyPersonality data set relied on users reporting their political views on their Facebook profile (linked to their real life identity). As such, this data may suffer from a social desirability bias; people with political views that could be viewed unfavourably by their peers might decline to reveal them. The anonymity of Reddit means that r/PoliticalCompassMemes users do not have to worry about their ideologies being linked to their true identity and can freely reveal their political beliefs.

Our data also features a more sophisticated response variable than utilized by Kosinski et al. Our measure of ideology has two components corresponding to a two-dimensional conception of ideology. It also allows for users to identify as centrists on one or both dimensions. This enables us to make novel insights regarding the robustness of predictive accuracy when we include ideological centrists and the conduciveness of digital footprints to the estimation of different facets of ideology. 

Finally, our data set has 91,000 observations\footnote{In practice, some observations are removed from the training/testing sets for various reasons. The full amount of observations used in training, validating and testing each model is reported alongside metrics of model performance in Section \ref{Results-model}.} and is thus substantially larger than Kosinski et al.'s which features 9,752 observations with political labels.

\section{Empirical Methodology}
\label{Methodology}

Firstly, we mapped the original flairs to {left, center, right} and removed centrists to perform binary classification on users' economic ideology. We also mapped flairs to {authoritarian, center, libertarian}, and removed centrists to perform binary classification on users' social ideology.  We refer to these classification problems as the economic problem and social problem respectively. Exact details on how we mapped from the set of nine classes to economic classes and social classes can be seen in Appendix \ref{Appendix-Recoding}.

The removal of centrist users can be justified on the basis that they may represent measurement noise. For instance, when we plot the data in singular value decomposition (SVD) component space, we typically see separations between right and left and (to a lesser extent) authoritarian and libertarian users. However, centrist users are randomly splattered throughout the SVD component space indicating that centrist users are not `true' centrists and may effectively be declining to reveal their true ideology. We illustrate this in Figure \ref{fig.svd_scatter} and provide further demonstration in Appendix \ref{Appendix-svd}. 

Further, the existence of `true' centrists is debated in the political science literature; consider Duverger's proclamation: ``in politics, the centre does not exist"   \citep{Brogan1955}. More concretely, recent work has tested competing hypotheses that attempt to explain why centrist self-identification arises. There is evidence that ``placing oneself on the center does not indicate that the individual is ‘ideologically centrist’"   \citep{Rodon2015} but rather that centrist self-identification predominantly arises from individuals having insufficient political knowledge to place themselves on a left right scale (the \textit{uninterested hypothesis}) or individual's self-identification reflecting decisions to vote for centrist parties or candidates rather than genuine ideological centrism (the \textit{party-component hypothesis}). The \textit{uninterested hypothesis} clearly does not apply to users in our sample who willingly identify themselves as centrists and have an interest in politics but the \textit{party-component hypothesis} (and the empirical evidence supporting it) provides a compelling case to disregard those who have labeled themselves as centrists in our initial models. 

Finally, removing centrists allows for easier comparisons with the work of Kosinski et al.   \citep{kosinski2013private} in which the US party affiliation of Facebook users (Democrat or Republican) is predicted, excluding all users that were not labelled as either. 

\begin{figure}[!htbp]
\centering
  \fbox{\includegraphics[width=14cm,height=8.75cm]{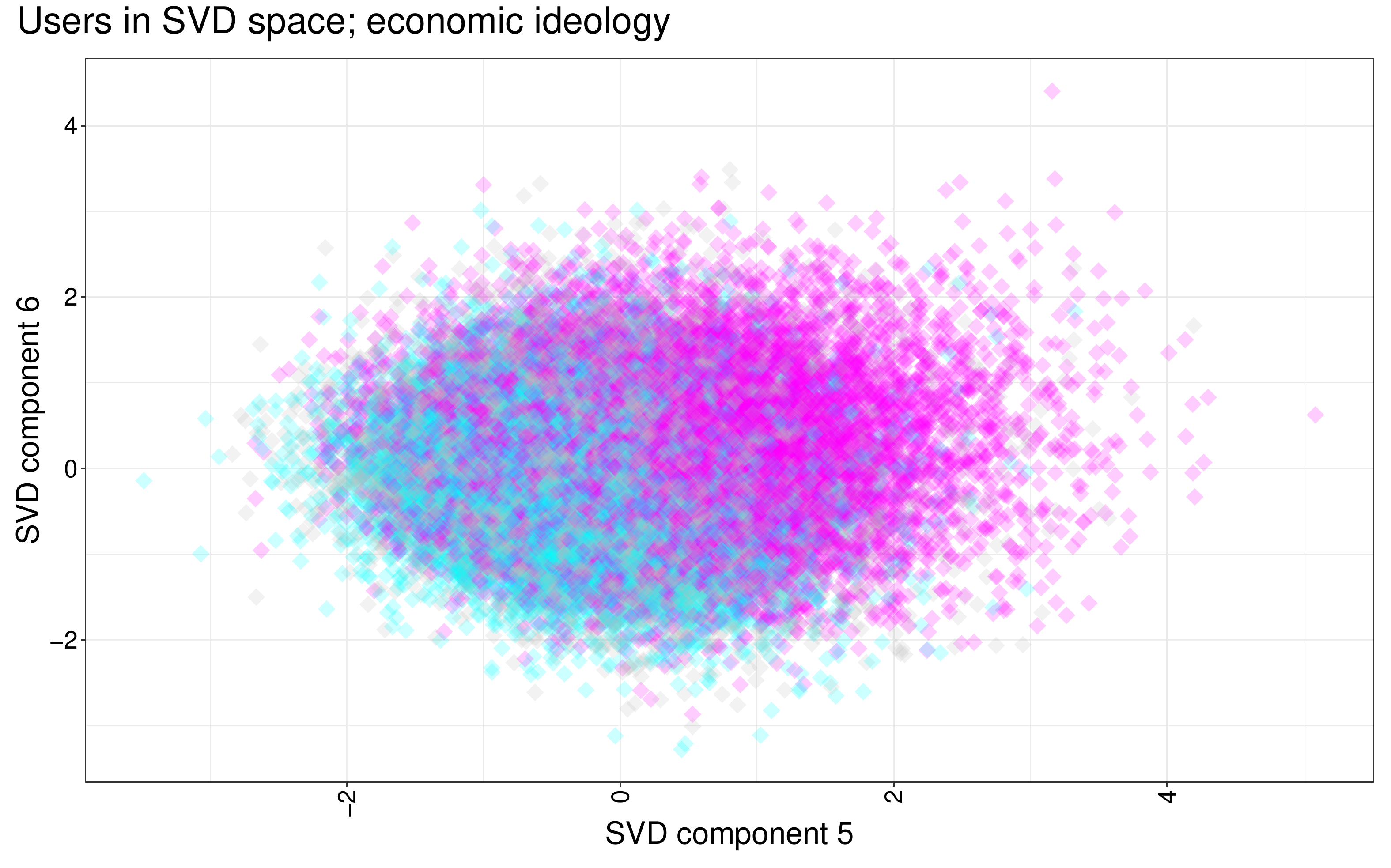}}%
  \caption{Leftwing (magenta), rightwing (cyan) and centrist (grey) users from  the training and validation sets in SVD component space \label{fig.svd_scatter}}
\end{figure}

It is not for us to determine the legitimacy of centrist self-identification, so whilst we present the results of binary classification, and note that they may be meaningful in their own right, we also estimated these models again (for both the economic and social problem), this time including centrists. Results from this multi-class classification problem demonstrate whether the results of the binary classifications are robust to the inclusion of so-called `centrists’. We then estimated models for the full nine-class classification problem in which we attempted to classify users to a class from the full set of categories that feature in our ideological labels. 

For each classification problem we ran models where we used as features A) the user-interaction matrix, B) features extracted from the textual content of comments and C) the union of these sets of features. 

Our data was split into training (64\% of total data), validation (16\%) and testing (20\%) sets. In all models, hyperparameters were chosen according to accuracy on the validation set. We then retrained the model on the training and validation data using the optimal hyperparameters. We report the accuracy and, where applicable, the (weighted) ROC-AUC of all models' predictions on our testing set and present these results as indicators of the models' predictive power over ideology. Model accuracy and ROC-AUC are detailed in Section \ref{Methodology-model-assessment}.

We expect accuracy to decrease as the number of classes in the response variable increases. As such, we assess the accuracy of all models relative to the baseline accuracy of a ZeroR classifier, which assigns all observations to the plurality class in the training set. For example, in the nine-class classification problem, the ZeroR classifier assigns all observations to ‘libleft’ ($\sim$ 20\% of all observations in sample). If digital footprints can predict ideology then our models will exhibit accuracy greater than the accuracy of the ZeroR classifier on the testing set. The more accurate our models are over the ZeroR baseline, the stronger the evidence that digital footprints can predict ideology.

The code for the Scikit-Learn implementation of all models for which results are reported in Section \ref{Results} can be found in the  \href{https://github.com/MichaelKitchener98/RedditIdeology-Honours2021/tree/main/Models}{models folder of the GitHub repository for this paper}\footnote{\url{https://github.com/MichaelKitchener98/RedditIdeology-Honours2021/tree/main/Models}}.

\subsection{Methodology: user-interaction matrix}
\label{Methodology-user-interaction}

When using the user-interaction matrix as our set of features we removed popular, explicitly political subreddits from the user-interaction matrix prior to running the models. We also trimmed our user-interaction matrix to ensure that every user in the data set had made at least 50 comments and every subreddit had been commented in at least 50 times as suggested by Kosinski et al.   \citep{Kosinski2016MiningBD}. This ensures that our user-interaction data better proxies digital footprints like search logs which are presumably expansive records. In many cases we `binarised' our data, i.e. for each cell $(i,\,j)$ of the user-interaction matrix, if $(i,\,j)$ is greater than 0 it was mapped to 1 in the binarised user-interaction matrix and 0 otherwise. The choice to binarise was made on the basis of validation set performance;  binarisation drastically increased performance for most models. 

In developing these models we followed the same broad approach as Kosinski et al.   \citep{kosinski2013private}:

\begin{enumerate}
    \item Reduce the dimensions of our predictors via a SVD of our data matrix. A smaller feature set is generally conducive to better performing models.
    \item Train a supervised learning algorithm on the SVD components of the user-interaction matrix to classify the ideology of a user.
\end{enumerate}

In step 2 we used the following learning models for multi-class classification: one-versus-rest (OVR) logistic regression (with and without an $\ell_1$ penalty), multinomial logistic regression (with and without an $\ell_1$ penalty), random forests, OVR random forests and AdaBoost. In the binary case we exclude OVR schemes from all models and multinomial logistic regression reduces to standard logistic regression. A brief account of each model and the SVD is given in Section \ref{Methodology-overview-stat-learning} excluding (OVR) random forests and AdaBoosting as these supervised learning algorithms do not feature in our best model for any modelling problem. An overview of (OVR) random forests and AdaBoosting can be found in Appendix \ref{Appendix-methodology}. We used the Scikit-Learn Python package   \citep{scikit-learn} to implement these models.

\subsection{Methodology: text}
\label{Methodology-text}

We utilised two approaches to extracting features from the textual content of users' comments. Firstly, we created features corresponding to the term frequency-inverse document frequency of different terms in the concatenation of users' comments (tf-idf), where documents are the sets of different users' comments. Secondly, we used a Word2Vec model   \citep{Mikolov2013, Mikolov2013a} that has been pre-trained on a Google News data set (`word2vec-google-news-300'\footnote{\url{https://drive.google.com/file/d/0B7XkCwpI5KDYNlNUTTlSS21pQmM/edit?resourcekey=0-wjGZdNAUop6WykTtMip30g}}) to convert the set of comments for each user into the simple average of the vector representations of their constituent words. 

We then  classified users' ideologies using a linear support vector classifier (SVC), applying a OVR scheme for multi-class problems. We restricted our choice of supervised learning algorithm to the linear SVC given that the strong performance of support vector machines in document classification is well established   \citep{Joachims1998} (we did not experiment with non-linear kernel functions owing to computational constraints). 

We ran the SVC using A) tf-idf features, B) average Word2Vec features and C) the union of these two sets of features. Some models used a subset of SVD components from a SVD of the original features instead of the original features themselves (this decision was made on the basis of validation set performance). We outline tf-idf, Word2Vec as well as the linear SVC model in Section \ref{Methodology-overview-stat-learning}.

\subsection{Methodology: combined}
\label{Methodology-combined}

These models use both the (often binarised) user-interaction matrix and the tf-idf vectors as features. We scaled each feature to have the same range since tf-idf terms and user-interaction records are not on the same scale.
We then took a SVD of the data if validation set accuracy suggested that this increased performance. We then fed this data into an OVR logistic regression model  with an $\ell_1$ penalty and an OVR linear SVC (or just a simple logistic regression and linear SVC when dealing with binary classification problems). 

We did not use the full range of models experimented with earlier
because OVR logistic regression with an $\ell_1$ penalty typically performed very well for models taking the user-interaction matrix as features. Further, the union of tf-idf and average Word2Vec features leads to (at best) marginally better results than models that just use tf-idf features.

\subsection{Methodology: user-interaction matrix features}
\label{Methodology-user-interaction-features}

Another facet to understanding the economic and social ideology is through understanding the features in terms of most relevant subreddits that help distinguish these ideologies. To identify this minimal subset of subreddits we employed the ‘Boruta’ \citep{kursa_feature_2010} feature selection algorithm on all of the subreddits. The algorithm creates copies of each individual feature in the dataset and randomises the values within each of the features. This randomly shuffled feature set is referred to as shadow features, and these are appended to the original data. A decision tree based classification algorithm (random forest) is then trained on the entire feature set (original features and shadow features). Features are then selected based on their importance. The threshold for the relevant feature importance is given by the maximum importance score of the shadow feature, thus ensuring that the feature contributes more information than a completely randomised feature column. 

We use the user-interaction matrix and the filters applied in the section \ref{Methodology-user-interaction} as input to our multi-labelled economic and social ideology ‘XGBoost’ classifiers \citep{chen_xgboost_2016}. The input user-interaction matrix is further filtered for relevant features. To identify the relevant features we train two feature selection models using Boruta (version 0.3) scikit-learn-contrib extension package \citep{Daniel2017} on the train dataset. One of the models is trained on a binarised dataset and the other is trained directly on the count dataset. The counts here are the number of user comments/posts in a given subreddit. The features are selected based on the cut-off threshold described earlier. We then train the multi-labelled economic and social ideology classifiers on the minimal feature set and compare the performance.

\subsection{Model assessment}
\label{Methodology-model-assessment}

\subsubsection{Accuracy}
\label{Methodology-model-assessment-accuracy}

The accuracy of a model on a test set indexed by $i=1,\dots,n$ is given by:

$$\text{accuracy} = \cfrac{1}{n} \sum^n_{i=1} I(y_i = \hat{y}_i) $$ 

Where $I(.)$ is an indicator function that returns $1$ if the argument inside it is true and $0$ otherwise, $y_i$ refers to the true ideology of user $i$, $\hat{y_i}$ refers to the predicted ideology of user $i$ given by our model. Consequently, accuracy represents the proportion of users in a test set whose ideology is correctly classified by our model. 

\subsubsection{ROC-AUC}
\label{Methodology-model-assessment-auc}

To calculate the OVR ROC curves for a multi-class classification model's predictions on our test set we need, for each class $k$ ($k = 0,\dots,K-1$) and for each member of our test set, $i$, estimates of the probability that person $i$'s ideology is $k$: $\hat{P}(y_i = k)$. Then, for a set of thresholds, $\tau \in [0,1]$, we assign observation $i$ to $k$ if $\hat{P}(y_i = k) > \tau$ and to `not $k$' otherwise. We then compute the sensitivity (true positive rate) and $1-$specificity (specificity is the true negative rate):

$$ \text{sensitivity} = \cfrac{\text{TP}}{\text{TP + FN}} =
\cfrac{\sum_{y_i = k}I(\hat{P}(y_i = k) > \tau)}
{\sum_{i=1}^n I(y_i = k)
}$$
$$ \text{specificity} = \cfrac{\text{TN}}{\text{FP + TN}} =
\cfrac{\sum_{y_i \not = k} I(\hat{P}(y_i \le \tau))}
{\sum_{i=1}^n I(y_i \not = k)}$$

We plot the values for sensitivity and $1-$specificity in $\langle \text{1-specificity, sensitivity} \rangle$ space for each $\tau$. The area under this curve is the ROC-AUC for class $k$; $\text{roc auc}_k$. We compute the AUC for each class and weight it by the total proportion of testing observations in class $k$; $p_k$. We sum these $K$ weighted, class specific ROC-AUC values to gain an overall testing ROC-AUC estimate for our model   \citep{989510}:

$$ \text{roc auc} = \sum^K_{k=1} \text{roc auc}_k \cdot p_k $$

The OVR ROC-AUC for class $k$ is equivalent to the probability that the classifier will assign a randomly selected observation from class $k$ a higher probability of membership in class $k$ than a randomly selected observation from some class other than $k$   \citep{989510}. Thus, it is essentially a measure of how well our classifier can distinguish between classes, weighted by class prevalence. This method of averaging OVR ROC-AUC scores is referred to as the `weighted' method in SciKit-Learn   \citep{scikit-learn}.

For binary classification problems we simply compute the probability that each observation belongs to one of the two classes (left/right or auth/lib) and compute the sensitivity and $1-\text{specificity}$ associated with varying thresholds to construct the ROC curve and therefore determine ROC-AUC.

ROC curves can only be constructed for models that can estimate the probability that an observation belongs to a class. As such, we omit the ROC-AUC metric from the evaluation of some models. Exactly how probability estimates are obtained from each model can be seen in the Scikit-Learn documentation   \citep{scikit-learn}.

\subsection{Overview of statistical learning approaches}
\label{Methodology-overview-stat-learning}

\subsubsection{Singular value decomposition}
\label{Methodology-svd}

Singular value decomposition (SVD) is a dimension reduction technique that allows us to replace our data matrix, $X_{(n \times p)}$, with a lower-dimensional representation: a matrix $\bar{X}_{(n \times q)}$ where $q < p$. The SVD computation finds the $p$-dimensional vector, $\vec{v_1}$, in feature space that minimises the projection error of data onto the vector. This is the first SVD component. We can then represent each data point by its projection onto this vector. The distribution of the projected values of data points has the largest variance of any possible projection. We then determine subsequent SVD components by finding the $p$-dimensional vectors that maximise the variance of projected data points subject to the constraint that the vectors are orthogonal to existing SVD components. We do this until we have $q$ vectors.

Mathematically, we are decomposing the data matrix, $X$, into the product of three matrices: 

$$ X_{(n \times p)} = U_{(n \times r)} \Sigma_{(r \times r)} V^{\intercal} _{(r \times p)}$$

Here, $U$ is an $n \times r$ matrix where the $n$ rows correspond to users and the $r$ columns correspond to `themes' in the subreddit interaction data; i.e. latent factors. Any particular cell $u_{(i,\,j)}$ intuitively represents how much user $i$ interacts with theme $j$. $\Sigma$ is a $(r \times r)$ diagonal matrix where  $\sigma_{(i,\,i)}$ represents the variance of theme $i$. $V$ is a $(p \times r)$ matrix that, when transposed, maps interaction with subreddits to themes, i.e. $v^{\intercal} _{(i,\,j)}$ represents how much each interaction with subreddit $j$ adds or detracts from the the score on theme $i$. $\vec{v_q}$ is the $q_{th}$ row of the matrix $V^{\intercal}$.

We chose to use the first 500 SVD components as features in the relevant models. In conjunction with trimming, this reduced the amount of features in the user-interaction matrix from over 190,000 to 500. This accounts for roughly $86\%$ of the variance of the original data. 

As noted by Kosinski et al., principal component analysis (PCA) is a less appropriate technique for the kind of problem we are dealing with. PCA requires that the data be centered which does not preserve sparsity and therefore undermines efficient computation   \citep{Kosinski2016MiningBD}.

\subsubsection{OVR logistic regression}
\label{Methodology-ovr-logreg}

We first describe a simple logistic regression (as used in our binary classification problems) and then explain how it can be extended to use in multi-class classification problems through a OVR scheme. 

A logistic regression models assumes that, when individual $i$ makes the binary decision $y_i \in \{ 0,\: 1\}$, the probability they choose $y_i=1$ is given by:

$$ P(y_i = 1|x_i;\,\beta) = \cfrac{e^{x_i^{\intercal} \beta}}{1+e^{x_i ^{\intercal} \beta}}$$

Where $x_i$ is a vector of predictors (often 500 SVD components in our case) with $x_{i, 0} = 1$ to allow for a constant term. $\beta = (\beta_0,  \beta_1, \dots, \beta_{500})^{\intercal}$ is a vector of coefficients to be estimated. Estimated model coefficients, $\hat{\beta}$, are obtained via maximum likelihood estimation (MLE), i.e. through solving the following maximization problem   \citep{10.5555/2834535} (where $i = 1,\dots,n$ now indexes observations in our training set):

$$ \hat{\beta} = \underset{\beta}{\mathrm{argmax}} \left\{ 
\frac{1}{n} \sum^n_{i=1} w_i \{ 
y_i \ln P(y_i = 1|x_i;\,\beta) + (1-y_i) \ln (1 - P(y_i = 1|x_i;\,\beta) )
\}
\right\}$$

$w_i$ is the weight for each observation. We try both a uniform weighting and balanced weighting: $w_i = \frac{n}{n_i}$  where $n_i$ refers to the number of observations in the same class as observation $i$. 

With our MLE estimate of $\beta$, $\hat{\beta}$, we can can estimate the probability that some person $j$, who does not feature in our training set, chooses $y_j=1$ as:

$$ \hat{P}(y_j = 1|x_i; \, \hat{\beta}) = \cfrac{e^{x_j^{\intercal} \hat{\beta}}}{1+e^{x_j ^{\intercal} \hat{\beta}}} $$

This works for binary dependent variables but we are often attempting to classify ideologies where our dependent variable, $y_i$, takes on one of the following forms:

\begin{itemize}
    \item $y_i \in \{\text{libleft, libright, libcenter, centrist, left, right, authright, authleft}\}$
    \item $y_i \in \{\text{left, center, right}\}$
    \item $y_i \in \{\text{authoritarian, center, libertarian}\}$
\end{itemize}

We can extend the binary model using a technique called OVR logistic regression which involves fitting a model for each class to predict the probability that someone is in that class versus not in that class.

Consider the economic ideology classification problem where $y_i \in \{\text{left, center, right}\}$. The OVR logistic regression approach to modelling this problem involves fitting three different logistic regression models ($\beta,\, \gamma,\text{ and } \delta$ are all separate coefficient vectors to be estimated, MLE estimates are denoted by $\hat{\beta},\, \hat{\gamma},\text{ and } \hat{\delta}$). 

\textbf{Model 1 - predict probability that individual is left versus not left:}

$$ \hat{P_1}(y_i = \text{left}|x_i;\,\hat{\beta}) = \cfrac{e^{x_j^{\intercal} \hat{\beta}}}{1+e^{x_j ^{\intercal} \hat{\beta}}} $$

\textbf{Model 2 - predict probability that individual is center versus not center:} 

$$ \hat{P_2}(y_i = \text{center}|x_i;\,\hat{\gamma}) = \cfrac{e^{x_j^{\intercal} \hat{\gamma}}}{1+e^{x_j ^{\intercal} \hat{\gamma}}} $$

\textbf{Model 3 - predict probability that individual is right versus not right:}

$$ \hat{P_3}(y_i = \text{right}|x_i;\,\hat{\delta}) = \cfrac{e^{x_j^{\intercal} \hat{\delta}}}{1+e^{x_j ^{\intercal} \hat{\delta}}} $$

We assign individual $j$ to the class for which the relevant OVR probability expression is largest. For example, we assign $j$ to left iff:

$$ \cfrac{e^{x_j^{\intercal} \hat{\beta}}}{1+e^{x_j ^{\intercal} \hat{\beta}}} > 
\cfrac{e^{x_j^{\intercal} \hat{\gamma}}}{1+e^{x_j ^{\intercal} \hat{\gamma}}}, \text{ and }
\cfrac{e^{x_j^{\intercal} \hat{\beta}}}{1+e^{x_j ^{\intercal} \hat{\beta}}} > 
\cfrac{e^{x_j^{\intercal} \hat{\delta}}}{1+e^{x_j ^{\intercal} \hat{\delta}}}$$

\subsection{OVR logistic regression with \texorpdfstring{$\ell_1$}{Lg} penalty}
\label{Methodology-ovr-logreg-l1}

To improve upon the basic logistic regression model, we can augment the log likelihood by subtracting (a proportion of) the $\ell_1$ norm of the coefficient vector from the log likelihood. Thus, we obtain coefficient estimates through solving the following maximization problem:

$$ \hat{\beta} = \underset{\beta}{\mathrm{argmax}} \left\{ 
\frac{1}{n} \sum^n_{i=1} w_i \{ 
y_i \ln P(y_i = 1|x_i;\,\beta) + (1-y_i) \ln (1 - P(y_i = 1|x_i;\,\beta) )
\} - \lambda ||\beta_{/0}||_1 \}
\right\}$$

Where $\lambda > 0$ and $||\beta_{/0}||_1$ refers to the $\ell_1$ norm of the $(p \times 1)$ vector $\beta_{/0}$, which is the vector of coefficients associated with all regressors excluding  the intercept term. i.e. $||\beta_{/0}||_1 = \sum^p_{j=1} |\beta_j|$.  

When coefficients deviate from zero in magnitude, $\lambda ||\beta_{/0}||_1$ increases, thus penalising the augmented log-likelihood. This has the effect of shrinking unimportant coefficients to zero and may improve predictive accuracy if the decrease in model variance more than offsets the increase in model bias   \citep{10.5555/2517747}. The lasso penalty can be viewed as the imposition of a constraint on the size of parameters onto the maximisation problem discussed in Section \ref{Methodology-ovr-logreg}. The parameter $\lambda$ is a tuning parameter that controls the strength of the regularization; a higher $\lambda$ implies stronger regularization i.e. a smaller budget for the maximum $\ell_1$ norm of the coefficient vector. 

This process provides a different set of coefficient estimates (with many possibly being equal to zero, effectively dropping irrelevant features from the model) but the process of prediction and implementation of the OVR scheme is the same as in Section \ref{Methodology-ovr-logreg}.

\subsubsection{Multinomial logistic regression}
\label{Methodology-multinomial}

Multinomial logistic regression is an extension of logistic regression that can naturally perform multi-class classification. When there are $K$ choices ($k = 0,\dots,K-1$), the probability that individual $i$ chooses choice $k$ is given by:

$$P(y_i = k |x_i,\; \beta_k) = \begin{cases}
\cfrac{1}{1+\sum^K_{k\prime=1} e^{x_i ^{\intercal} \beta_{k\prime}}} & k = 0,\\
\cfrac{e^{x_i ^{\intercal} \beta_k}}{1+\sum^K_{k\prime=1} e^{x_i ^{\intercal} \beta_{k\prime}}} & k \not = 0 
\end{cases}$$

Thus, prediction requires the estimation of a separate coefficient vector $\beta_k$ for each class. By convention, $\beta_0$ is set to $\vec{0}$ since we cannot uniquely identify any particular $\beta_k$, we can only identify $(\beta_k - \beta_0)$   \citep{10.5555/2834535}. 

Coefficient vector estimates, $\hat{\beta}_1,\dots,\hat{\beta}_{K-1}$,  are found by solving the following maximization problem:

$$ \hat{\beta}_1,\dots,\hat{\beta}_{K-1} = \underset{\beta_1,\dots,\beta_{K-1}}{\mathrm{argmax}} \left\{ \frac{1}{n}\sum^n_{i=1} w_i \sum^K_{k=1} I(y_i = k) \ln P(y_i = k | \beta_k) \right\} $$ 

This allows us to compute the predicted probability that a new observation, $j$, belongs to any particular class like so: 

$$\hat{P}(y_i = k | x_i;\;\hat{\beta}_k) = \begin{cases}
\cfrac{1}{1+\sum^K_{k\prime=1} e^{x_i ^{\intercal} \hat{\beta}_{k\prime}}} & k = 0,\\
\cfrac{e^{x_i ^{\intercal} \hat{\beta}_k}}{1+\sum^K_{k\prime=1} e^{x_i ^{\intercal} \hat{\beta}_{k\prime}}} & k \not = 0 
\end{cases}$$

We assign $j$ to $\underset{k}{\mathrm{argmax}}\, \hat{P}(y_j = k|\hat{\beta}_k)$ to obtain a categorical prediction.

\subsubsection{Multinomial logistic regression with \texorpdfstring{$\ell_1$}{Lg} penalty}
\label{Methodology-multinomial-l1}

The $\ell_1$ penalty described in Section \ref{Methodology-ovr-logreg-l1} can also be fruitfully applied to multinomial logistic regression. Thus, we are choosing $\hat{\beta}_1,\dots,\hat{\beta}_{K-1}$ to maximise   \citep{10.5555/2834535}:

$$ \hat{\beta}_1,\dots,\hat{\beta}_{K-1} = \underset{\beta_1,\dots,\beta_{K-1}}{\mathrm{argmax}} \left\{ \frac{1}{n}\sum^n_{i=1} w_i \sum^K_{k=1} I(y_i = k) \ln P(y_i = k | \beta_k)  - \lambda \sum^K_{k=1} ||\beta_{k,/0}||_1 \right\}$$ 

Where $\beta_{k,/0}$ refers to the coefficient vector $\beta_k$ without the intercept coefficient. 

This has the same effect as described in Section \ref{Methodology-ovr-logreg-l1}   \citep{10.5555/2834535}. The rest of the classification procedure after model estimation is the same as described in Section \ref{Methodology-multinomial}.

\subsubsection{OVR linear support vector classifier}
\label{Methodology-svc}

We describe a linear support vector classifier (SVC) for a binary classification problem. This is extended to multi-class classification via a OVR scheme that determines an observation's class on the basis of its distance between several, class specific separating hyperplanes. 

The SVC estimates a linear decision boundary in feature space by finding the hyperplane that maximises the margin (euclidean distance between the hyperplane and the observations closest to the hyperplane from either class) subject to a certain tolerance for training observations that violate the margin (i.e. that are on the `wrong' side of the margin). 

The coefficients for this hyperplane, $\hat{\beta}= (\hat{\beta}_0, \hat{\beta}_{/0}^{\intercal})^{\intercal}$ are found by solving the following problem   \citep{10.5555/2517747}:

$$
\hat{\beta} = 
\underset{\beta}{\mathrm{argmin}} =
\cfrac{1}{2} \beta_{/0}^\intercal \beta_{/0} + C
\sum_{i=1}^n \max(0,\, y_i(\beta_{/0}^\intercal x_i + \beta_0))$$

Here $x_i$ does not include a constant first element equal to one. $C$ is a constant that is \textit{inversely} proportional to the strength of regularization\footnote{In some formulations of the linear SVC maximization problem C refers to the regularization parameter itself but we follow the formulation given in the Scikit-Learn documentation.}. Higher values of $C$ will correspond to a stricter margin with less tolerance for violation.

We can predict the class of an observation, $j$, depending on which side of the estimated separating hyperplane it falls, i.e depending on the sign of $x_j^{\intercal}\hat{\beta}$.

\subsubsection{Term frequency - inverse document frequency}
\label{Methodology-tfidf}

The first step in this process of feature extraction is to `clean' the text. We remove all html links, punctuation, non ascii symbols (i.e. emojis) and single letter terms, convert all terms to lower case, and finally stem all terms. 

Then, for each word, $w$, we compute the `term frequency' for user $i$:

$$TF_{w,i} = \cfrac{n_{w,i}}{\sum_{k=1}^{K} n_{k,i} }$$

Where $n_{w,i}$ refers to how many times word $w$ appears in user $i$'s concatenated comments and $k=1,\dots,K$ indexes all words appearing in our vocabulary of terms (which is defined by the terms in the total set of all comments).

We then compute the inverse document frequency (IDF) of each word\footnote{This specific form is implemented by our models and differs slightly from standard textbook definitions of inverse document frequency.}:

$$IDF_w = \ln \left( \cfrac{D+1}{D_w+1}\right)  + 1$$

Where $D$ refers to the total number of users whose comments we have recorded (total amount of documents) and $D_w$ is the amount of concatenated sets of comments containing word $w$. 

Consequently, $TF_{w,i}\cdot IDF_w$ measures how much user $i$ uses word $w$ relative to other users. We construct the inverse document frequency for every word in our training vocabulary that has not been excluded by some criteria and then create tf-idf scores for each user (both in the training and testing set) for all of these terms. We then normalise tf-idf vectors for each user by their euclidean norm. 

Several hyperparameters impact the set of terms for which we compute tf-idfs, namely the maximum and minimum frequency of words which we allow in our vocabulary. Words that occur in a greater proportion of documents than our maximum frequency or in a smaller proportion of documents than our minimum frequency are disregarded. We also specify the maximum amount of words in the vocabulary, where words are ranked according to their term frequency across the corpus.

\subsubsection{Word2Vec}
\label{Methodology-embed}

Word2Vec is an unsupervised learning algorithm that takes a corpus of text and `learns' to represent each word as a vector. We use embedding from a model that has been pre-trained  on Google News data. This model contains 300-dimensional vectors for around 100,000,000,000 words and does not require that we convert our text to lower case or remove punctuation. Using a pre-trained model allows us to avoid training our own model which is time consuming and may be inaccurate due to our relatively small corpus.

The workings of the Word2Vec algorithm are somewhat complex so for the sake of brevity we refer the reader to    \citep{Mikolov2013} and    \citep{Mikolov2013a}.

\section{Results}
\label{Results}

\subsection{Model results}
\label{Results-model}

The accuracy and ROC-AUC (where applicable) of every model estimated for all binary and three-class models can be found in Table \ref{table.results_1}

\begin{table}[!htbp] 
  \centering
  \caption{Binary and multi-class classification of economic and social ideology \\ 
\scriptsize  N refers to the total amount of samples divided between training, validation and testing sets. In the multi-class problem AUC refers to weighted ROC-AUC}
      \scalebox{0.70}{
    \begin{tabular}{lllllllll}
    \makebox[0pt][l]{\textbf{Binary economic classification}} &       &       &       &       & \makebox[0pt][l]{\textbf{Binary social classification}} &       &       &  \\
    Model & Accuracy & AUC   & N     &       & Model & Accuracy & AUC   & N \\
\cmidrule{1-4}\cmidrule{6-9}    \makebox[0pt][l]{\textit{Features: user-interaction matrix}} &       &       &       &       &       &       &       &  \\
\cmidrule{1-4}\cmidrule{6-9}    ZeroR & 56.06\% &       & 52,881 &       & ZeroR & 78.99\% &       & 54,151 \\
    Random forest & 73.38\% & 81.29\% & 52,881 &       & Random forest & 80.47\% & 72.63\% & 54,151 \\
    AdaBoost & 79.07\% & 87.76\% & 52,881 &       & AdaBoost & 81.29\% & 73.98\% & 54,151 \\
    Logit ($\ell_1$) & 82.37\% & 90.63\% & 52,881 &       & Logit($\ell_1$) & 82.00\% & 78.66\% & 54,151 \\
    Logit  & 82.38\% & 90.63\% & 52,881 &       & Logit  & 82.02\% & 78.66\% & 54151 \\
          &       &       &       &       &       &       &       &  \\
    \makebox[0pt][l]{\textit{Features: textual features}} &       &       &       &       &       &       &       &  \\
\cmidrule{1-4}\cmidrule{6-9}    ZeroR & 54.07\% &       & 57,045 &       & ZeroR & 76.62\% &       & 59,055 \\
    Linear SVC; w2v & 66.68\% &       & 57,045 &       & Linear SVC; w2v & 76.62\% &       & 59,055 \\
    Linear SVC; tf-idf & 72.57\% &       & 57,045 &       & Linear SVC; combined & 79.94\% &       & 59,055 \\
    Linear SVC; combined & 73.21\% &       & 57,045 &       & Linear SVC; tf-idf & 79.98\% &       & 59,055 \\
          &       &       &       &       &       &       &       &  \\
    \makebox[0pt][l]{\textit{Features: combined}} &       &       &       &       &       &       &       &  \\
\cmidrule{1-4}\cmidrule{6-9}    ZeroR & 56.06\% &       & 51,199 &       & ZeroR & 79.42\% &       & 52,399 \\
    Logit ($\ell_1$) & 82.29\% & 90.38\% & 51,199 &       & Logit ($\ell_1$) & 82.69\% & 79.12\% & 52,399 \\
    Linear SVC & 82.83\% &       & 51,199 &       & Linear SVC & 83.09\% &       & 52,399 \\

  & &  &  & & &  &  & \\
 
    \makebox[0pt][l]{\textbf{Multi-class economic classification}} &       &       &       &       & \makebox[0pt][l]{\textbf{Multi-class social classification}} &       &       &  \\
\cmidrule{1-4}\cmidrule{6-9}    \makebox[0pt][l]{\textit{Features: user int}} &       &       &       &       &       &       &       &  \\
\cmidrule{1-4}\cmidrule{6-9}    ZeroR & 36.63\% &       & 80,961 &       & Multinomial logit ($\ell_1$) & 48.16\% & 66.68\% & 80,961 \\
    OVR Random Forest & 51.10\% & 69.68\% & 80,961 &       & ZeroR & 52.84\% &       & 80,961 \\
    Random forest & 51.13\% & 70.79\% & 80,961 &       & Random forest & 53.93\% & 62.58\% & 80,961 \\
    AdaBoost & 53.69\% & 72.03\% & 80,961 &       & OVR Random Forest & 53.98\% & 63.72\% & 80,961 \\
    OVR Logit  & 57.36\% & 76.43\% & 80,961 &       & AdaBoost & 54.00\% & 63.21\% & 80,961 \\
    Multinomial logit ($\ell_1$) & 57.50\% & 76.59\% & 80,961 &       & Multinomial logit & 55.82\% & 67.35\% & 80,961 \\
    Multinomial logit & 57.52\% & 76.54\% & 80,961 &       & OVR Logit  & 55.94\% & 67.31\% & 80,961 \\
    OVR Logit ($\ell_1$) & 57.65\% & 76.51\% & 80,961 &       & OVR Logit ($\ell_1$) & 55.94\% & 67.31\% & 80,961 \\
          &       &       &       &       &       &       &       &  \\
    \makebox[0pt][l]{\textit{Features: textual features}} &       &       &       &       &       &       &       &  \\
\cmidrule{1-4}\cmidrule{6-9}    ZeroR & 35.07\% &       & 87,547 &       & ZeroR & 51.40\% &       & 87,547 \\
    Linear SVC; w2v & 41.56\% &       & 87,547 &       & Linear SVC; w2v & 51.91\% &       & 87,547 \\
    Linear SVC; tf-idf & 48.94\% &       & 87,547 &       & Linear SVC; combined & 53.14\% &       & 87,547 \\
    Linear SVC; combined & 49.00\% &       & 87,547 &       & Linear SVC; tf-idf & 53.71\% &       & 87,547 \\
          &       &       &       &       &       &       &       &  \\
    \makebox[0pt][l]{\textit{Features: combined}} &       &       &       &       &       &       &       &  \\
\cmidrule{1-4}\cmidrule{6-9}    ZeroR & 36.83\% &       & 78,348 &       & ZeroR & 53.22\% &       & 78,348 \\
    OVR Logit ($\ell_1$) & 57.31\% & 76.46\% & 78,348 &       & OVR Logit ($\ell_1$) & 56.28\% & 67.17\% & 78,348 \\
    Linear SVC & 57.63\% &       & 78,348 &       & Linear SVC & 56.56\% &       & 78,348 \\
    
    &       &       &       &       &       &       &       &  \\
    \makebox[0pt][l]{ \textit{Features: feature selection}} &       &       &       &       &       &       &       &  \\
    \cmidrule{1-4}\cmidrule{6-9} XGBoost (Binary) & 55.65\% & 74.98\% & 80,961 &       & XGBoost (Binary) & 46.33\% &    66.04\%   & 80,961 \\
    XGBoost (Count) & 54.92\% & 74.77\% & 80,961 &       & XGBoost (Count) & 46.05\% &    65.70\%   & 80,961 \\
    \end{tabular}}
  \label{table.results_1}%
\end{table}%

In the binary classification of economic ideology ($y_i \in \{ \text{left, right} \}$) using only the user-interaction matrix as features, our best model\footnote{Best model refers to the model with the highest testing accuracy.}, logistic regression's, predictions on the test set achieved an accuracy of 82.38 \% (against a ZeroR baseline of 56.06 \%) and an ROC-AUC of 90.63\%. This reinforces Kosinski et al’s findings. Kosinski et al. were able to predict whether a Facebook user was a Democrat or Republican based off their Facebook Likes   \citep{kosinski2013private} (noting that they do not omit Facebook Likes for explicitly political pages) with an ROC-AUC of 85\%.   

In contrast, our best model in the binary classification of social ideology ($y_i \in \{ \text{lib, auth}\}$) using only the user-interaction matrix, logistic regression, barely improved upon the ZeroR baseline accuracy of 78.99\%, achieving an accuracy of 82.02\% on testing data. The model achieved an ROC-AUC of 78.66\% on the test set\footnote{The high ROC-AUC figure here cannot be trusted as the response variable is highly imbalanced as indicated by the ZeroR classification accuracy.}.

Textual features were worse performers in the binary classification problems. In predicting economic ideology, the linear SVC applied to the union of tf-idf features and average Word2Vec features (which was the best model in this task) achieved an accuracy of 73.21\% relative to the 54.07\% ZeroR baseline. For social ideology, the best model, a linear SVC using just tf-idf features, achieved an accuracy of 79.98\% against a 76.62\%  baseline\footnote{ZeroR baselines slightly differ between models due to different splits.}. Interestingly, this illustrates that the difficulty of predicting social ideology relative to economic ideology is robust to the type of digital record employed. 

The models that utilised both the user-interaction matrix and textual footprints (tf-idf features) did not substantially improve upon the performance of the models that solely used the user-interaction matrix as features. 

We now report the performance of models in the economic and social ideology classification tasks when we did not omit centrist observations, i.e. when we classify $y_i \in \{ \text{left, center, right} \}$ and $y_i \in \{ \text{lib, center, auth}\}$. Using just the user-interaction matrix, in the economic classification task our best model, OVR logistic regression with an $\ell_1$ penalty, achieved an accuracy of 57.65\% against a ZeroR baseline of 36.63\% and an ROC-AUC of 76.51\%. For the social problem, OVR logistic regression marginally surpassed baseline performance, achieving an accuracy of 55.94\% (baseline: 52.84\%) and an ROC-AUC of 67.31\%.

The models using textual features again perform worse than those using the user-interaction matrix. The best economic classifier using textual features (linear SVC using a union of tf-idf features and average Word2Vec features) still performs substantially better than baseline benchmarks (accuracy: 49.00\% vs 35.07\% ZeroR accuracy). The best social classification model (linear SVC using just tf-idf features) achieves an accuracy of 53.71\% against a 51.40\% baseline. This indicates that the classification of economic ideology is robust to the inclusion of centrists (i.e. we can still classify users’ ideology  at an accuracy substantially above baseline), albeit with significantly diminished accuracy. We also see that the stronger performance of the user-interaction matrix relative to textual features is robust to the inclusion of centrists. 

We suspect the primary reason for the comparatively poor performance of textual features is the limited variation in language use in our sample. The r/PoliticalCompassMemes subreddit is an online subculture with its own vernacular. For instance, many users of all ideologies use jargon such as `based’ and `cringe’ to indicate appreciation or dissatisfaction. We scraped comments from subreddits outside of r/PoliticalCompassMemes to try and maximise variation in word use but the use of the vernacular likely persists in other subreddits. Further, though we are unable to rigorously validate this claim, most users are likely young; many users in our sample comment in subreddits indicative of age, i.e. r/teenagers and appear to be located in the United States based on the prevalence of discussion topics relevant to US politics.

\begin{figure}[!htbp]
\centering
  \fbox{\includegraphics[width=14cm,height=8.75cm]{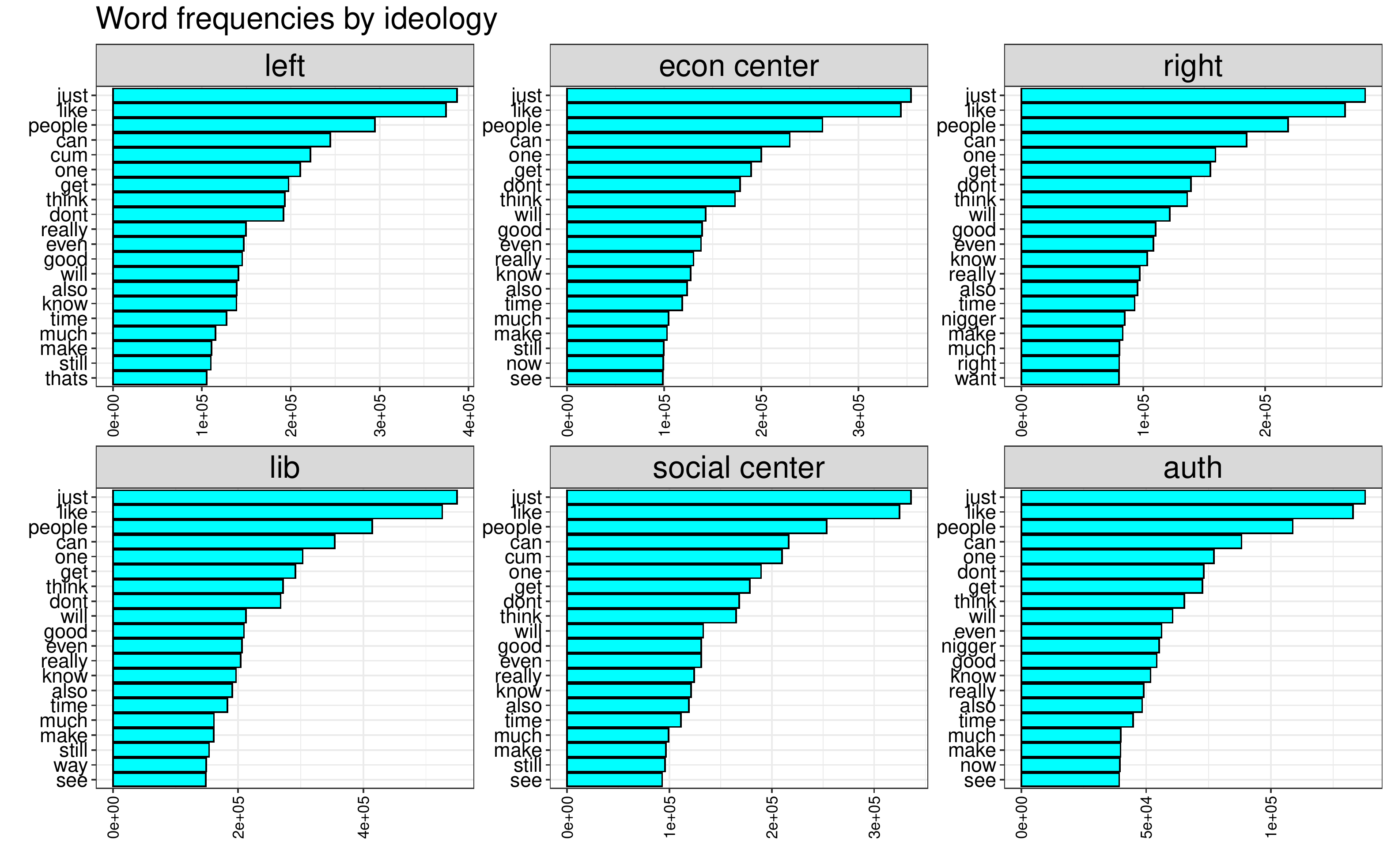}}
  \caption{Word frequencies in comments by ideology \label{fig.word_freq}}
\end{figure}

This effectively restricts the variation in vocabulary used by r/PoliticalCompassMemes commentors making it harder to link variation in language to ideology. The lack of variation in language is shown in Figure \ref{fig.word_freq} (the frequency of the n-word and `cum' is due to a handful of `troll' users who have made many comments which repeat these words thousands of times).

\begin{figure}[!htbp]
\centering
  \fbox{\includegraphics[width=14cm,height=14cm]{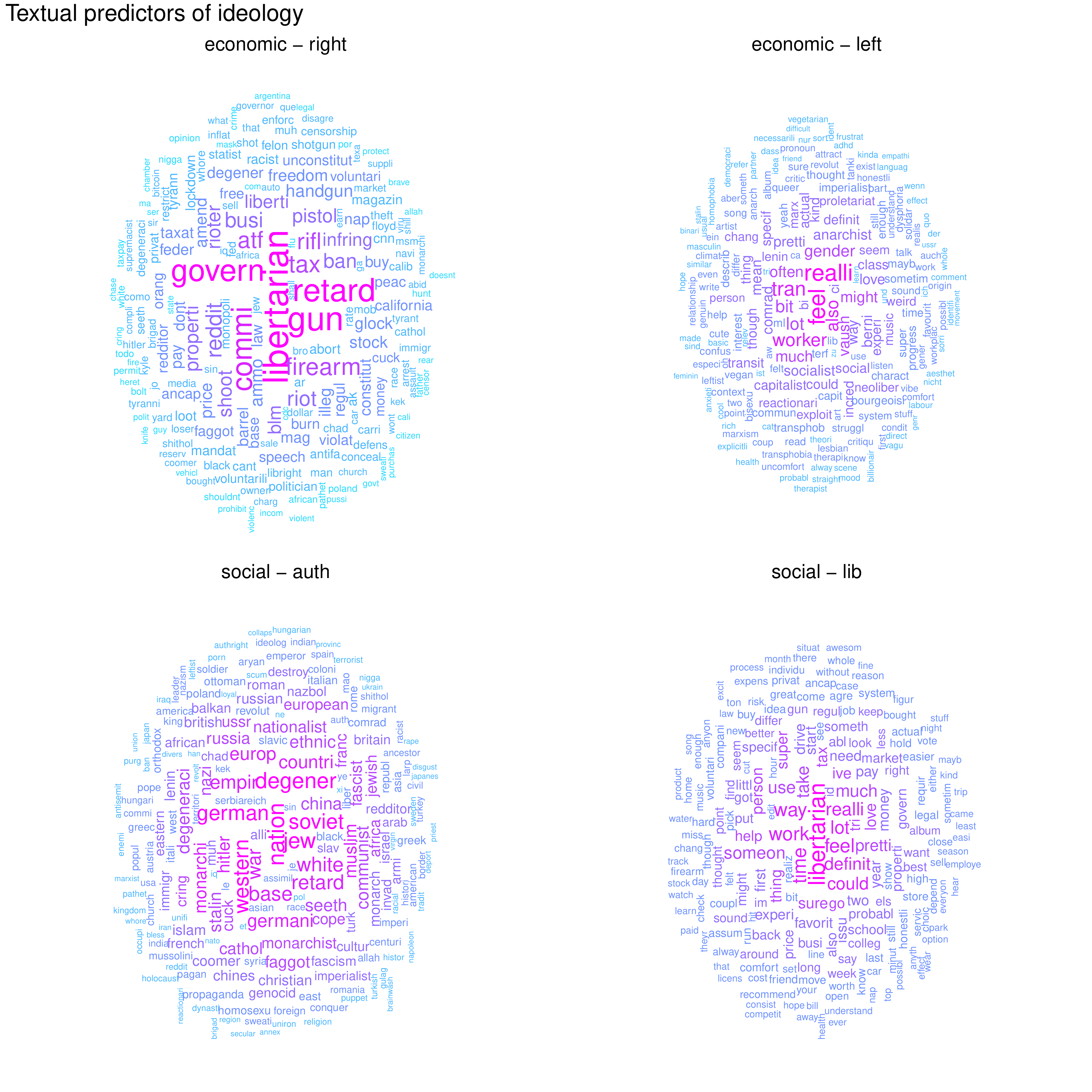}}%
  \caption{Economic ideology was recoded from \{left, center, right\} to \{-1, 0, 1\} and social from \{lib, center, auth\} to \{-1, 0, 1\}. We then computed the correlation of the tf-idf scores for each term with the quantitative ideology measure. Size of terms in wordclouds is proportional to correlation with the relevant ideology 
  \label{fig.wordcloud}}
\end{figure}

Still, variation in language is associated with ideology and textual features remain useful. To provide some insight into which facets of language are most associated with ideology we display the words most strongly correlated with economic and social ideologies in Figure \ref{fig.wordcloud}.

Again, in the multi-class classification problems, the union of the user-interaction matrix and textual features does not lead to models with substantively better performance on unseen data. We suspect that this is due to colinearity between the user-interaction features and textual features. In the same way that r/PoliticalCompassMemes has its own vernacular, interaction with  subreddits is likely associated with particular language usage. If a user has made comments in the r/Vegetarian subreddit then they have likely used language strongly associated with  vegetarianism, i.e. `vegan’, `meat’, `dairy’, increasing the value of textual features associated with use of these words. Thus, engagement with subreddits and language use likely go hand in hand in many cases and there is little information gained from textual features when we have already employed records of subreddit interaction. 

The final set of models predict which of the full nine classes each user belongs to, i.e. $y_i \in \{ \text{centrist, left, right, libright, libleft, libcenter, authright, authleft, authcenter}\}$. Understandably, these models feature the lowest accuracy due to the abundance of possible classes. Nevertheless, the best model based on the user-interaction matrix (multinomial logistic regression) was able to substantially improve upon baseline accuracy achieving an accuracy of 34.82\% against the 20.81\% ZeroR baseline and an AUC of 74.61\%. The best nine-class classification model using solely textual data (linear SVC on the union of tf-idf features and average Word2Vec features) performs worse than the best model using solely interaction data, achieving an accuracy of 24.4\% against the 19.84\% ZeroR baseline. The union of textual features with the user-interaction matrix does not substantially improve performance in the nine-class classification problem. Table \ref{table.results_2}, which provides the full set of results for all models in the nine-class problem, has been relegated to Appendix \ref{Appendix-nineclass} as these results are somewhat ancillary to our other findings.

The models that employ feature selection perform similar to or slightly better than the tree based algorithms (Random forest and AdaBoost). The AUC for binary based and count based feature selection models for economic ideology prediction is 74.98\% and 74.77\%, and the AUC for social ideology prediction is 66.04\% and 65.70\% respectively. The number of features used for count based feature selection for economic ideology prediction is 239, and binary based feature selection is 415, and 99.2\% of the features in the count based approach also appear in binary based feature selection. A similar pattern is observed in the social ideology prediction task - the number of features used for count based feature selection is 178, and all of these features except one are also present in the set of 357 features seen in the binary based feature selection model. Thus, we see the density of user-subreddit interaction to be an informative signal for ideology prediction. The selected feature list is provided in the Appendix \ref{Appendix-feat-selection}.

\subsection{Restricted Sampling Study}

To further interrogate the informational potential of user-subreddit interaction patterns for ideological inference, we undertake a restricted sampling exercise, in the spirit of \cite{youyou2015computer} (refer Figure 2, p.1038 in the reference). In the main modelling exercises reported so far, we employ the full dataset of users' interaction patterns on subreddits ($N > 50,000$). However, a natural question arises, `How little information does one need about a user's subreddit interaction activity to accurately infer their political ideology?' We have in mind contexts where research projects or applications might have access only to a limited dataset, or require relatively accurate inference at fast time-scales.

To explore this question, and working with the binary economic and social classification task, we first identify and fix a sample of roughly 7,000 users who have engaged in at least 200 unique subreddits. Whilst these are clearly `power' users on Reddit, it ensures that the user population is constant, with no user dropping out due to insufficient unique subreddit interactions as sampling sizes are increased. We then construct a series of user-subreddit interaction dataset built from a random sample of these users' subreddit interactions, starting with randomly sampling 10 unique subreddits only, through to 25, 50, 100 and 200. The sampled subreddits need not be the same for each user. Sampling is conducted independently at the user level. In effect, we aim to mimic the idea that a researcher or user has only a very limited capacity to observe a given Reddit user. In such a restricted informational environment, how well can a model infer political ideology? We fit a Logistic model to each sampling dataset, being the (equal) best performing model across both economic- and social- binary classification on the full sample.

\begin{figure}
    \centering
    \includegraphics[width=0.7\textwidth]{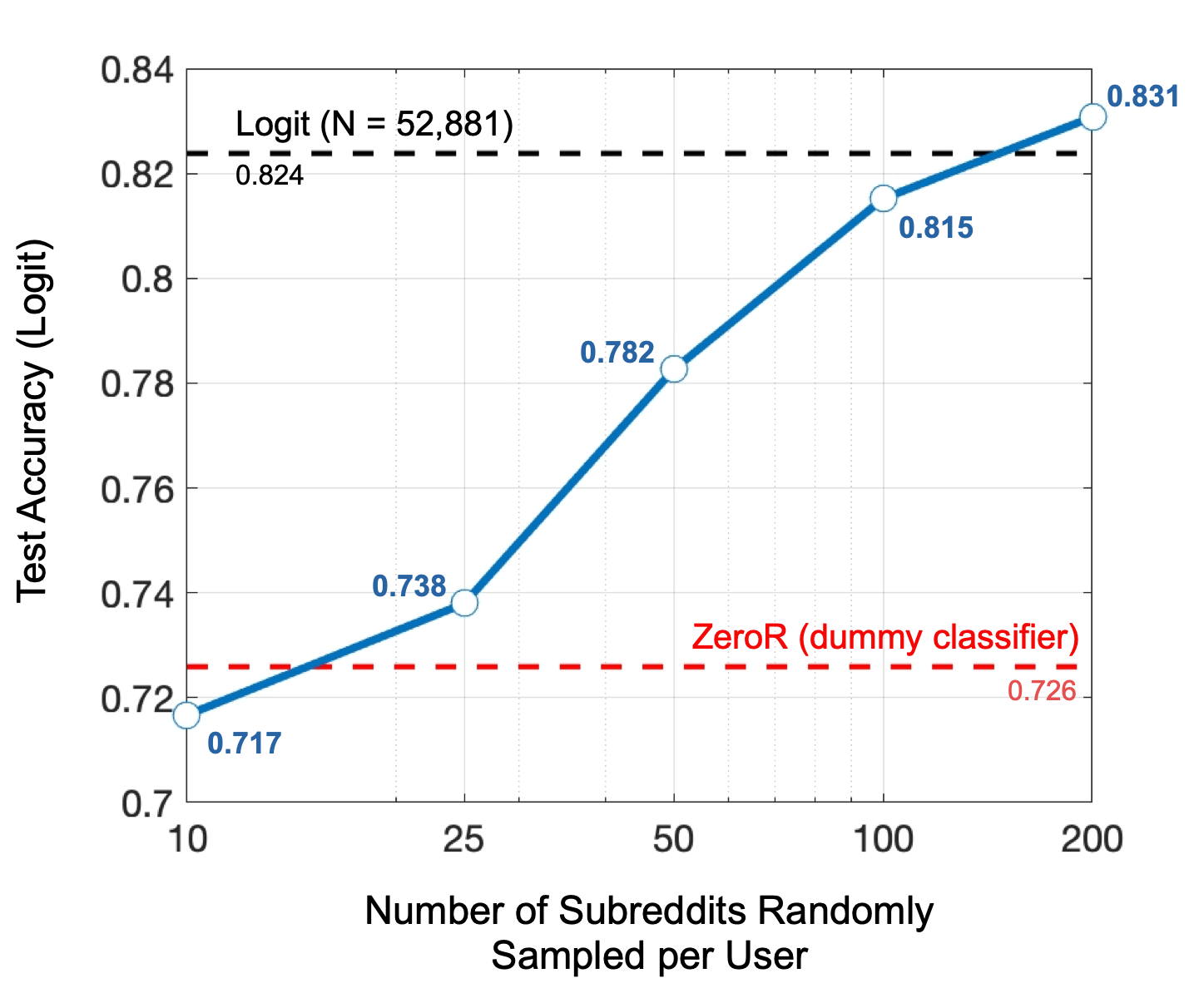}
    \caption{Restricted sampling study results showing model accuracy improvement against the ZeroR and best performing full-sample model benchmarks.}
    \label{fig:sampling}
\end{figure}

Figure~\ref{fig:sampling} presents model accuracy results for the binary economic classification task on the test dataset (train 5,644 users, test 1,412 users). The results are stark and mimic strongly the findings of \cite{youyou2015computer} (Fig.~2): whilst 10 unique subreddits carries insufficient information to infer binary economic ideology, this picture changes rapidly, 50 subreddits yields an accuracy gain of over 5\%, with the threshold of 80\% accuracy passed at around 80 subreddits. By 100 subreddits, accuracy of 81.5\% is close to the full sample accuracy of 82.4\%, across all available subreddit interactions for over 54,000 users. Whilst we do not, as in the reference, have the ability in this study to compare the computational inference accuracy with that of individuals in a user's social network, it is clear that even a few dozen subreddit interactions carries enough information to start identifying political ideology. A caveat to this result is that lower intensity users (e.g. those with less than 200 unique subreddit interactions) may not reveal as potent information in their interaction pattern as these power users, though we suspect the result will change little for such users, given the randomised sampling procedure. Subreddits included in a users' interaction pattern in this sampling study were equally chosen if the user had been in the subreddit for 1 day or 1000 days.

We conducted the same exercise for the binary social ideology task, but results were predictably weak. With little head-room to work with (ZeroR accuracy of 78.99\% to Logit accuracy of 82.02\%) in the full sample study, little discernable pattern was identified in the accuracy of Logit models trained on the restricted sampling interaction patterns.

\subsection{Textual Insights}
\label{Results-viz}

\begin{figure}[!htbp]
\centering
  \fbox{\includegraphics[width=14cm,height=17.5cm]{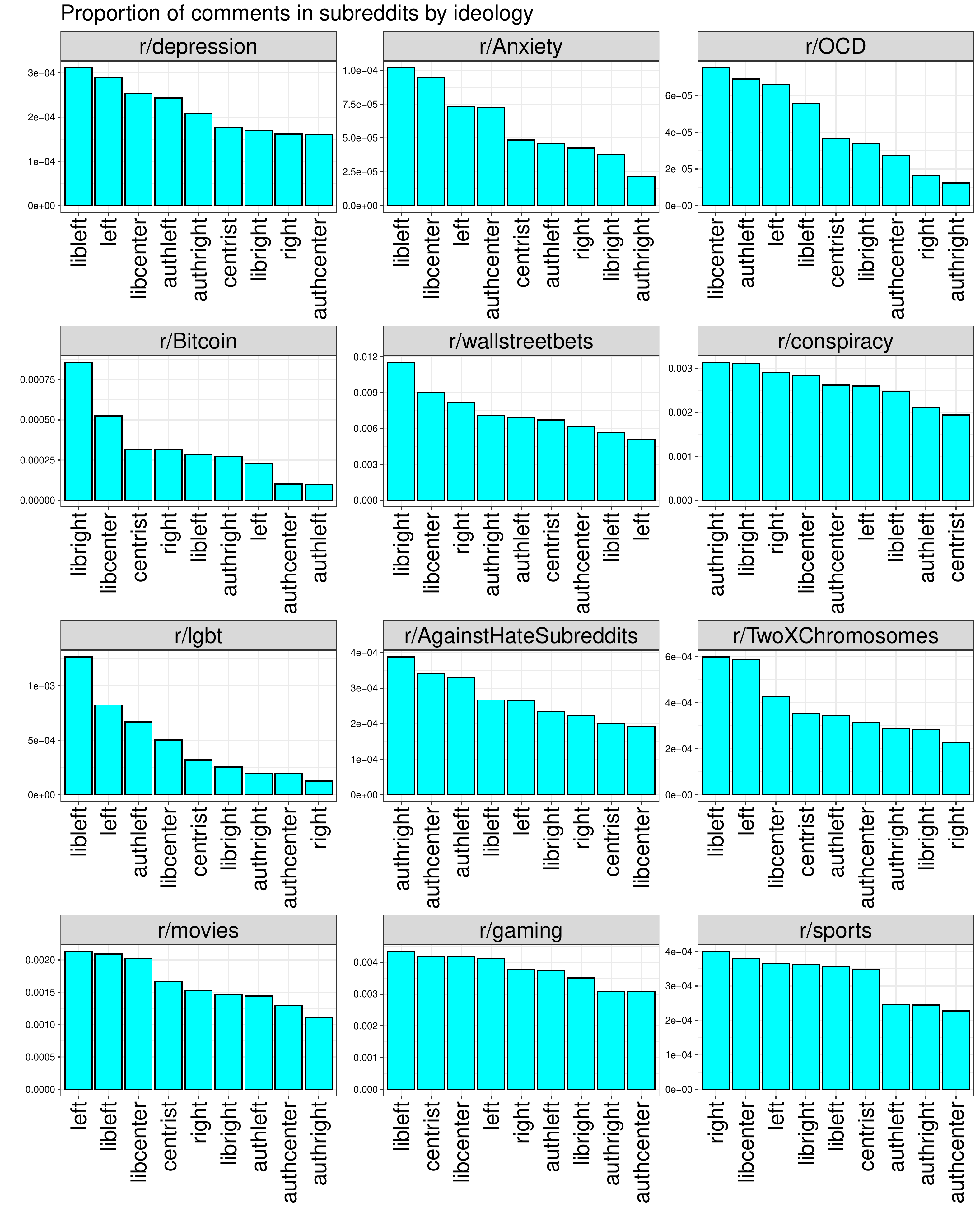}}%
  \caption{This figure shows the proportion of total comments from users of each ideology in a range of subreddits  \label{fig.int_eda}}
\end{figure}

In Figure \ref{fig.int_eda} we display the proportion of total comments from each ideology in  particular subreddits. This is intended to illustrate the variation in subreddit interaction between users of different ideologies and how Reddit data can illuminate psycho-social dynamics. From  the first row, one could infer that those with more left-leaning economic views are more comfortable discussing any mental illnesses they are facing. This also illustrates how digital footprints related to mental health (i.e. search engine queries relating to a condition) are the type of digital footprint that could provide information on a person's political beliefs despite the fact there is no obvious connection between mental health and political ideology.   

The second row of Figure \ref{fig.int_eda} illustrates how libertarian users of any economic persuasion (but especially those with right wing economic views) interact the most with r/Bitcoin. This makes sense as Bitcoin is a decentralised digital currency that can be used anonymously; which would naturally appeal to those who are skeptical of state power and value political freedom highly. We also see that economically right wing users interact most with the popular r/wallstreetbets subreddit which is focused on risky trading strategies in financial markets. Economically right wing users also interact more with the r/conspiracy subreddit. 

The third row of Figure \ref{fig.int_eda} shows these results for several subreddits that common wisdom dictates would be favoured by users with progressive views. r/lgbt is a subreddit for people with sexual orientations other than heterosexual, r/AgainstHateSubreddits is a forum focused on identifying and denouncing other subreddits perceived to have engaged in prejudiced behaviour and r/TwoXChromosomes is a subreddit for women (studies show women tend to lean further left). Variation in the amount of interaction with these subreddits along ideological lines allows for digital footprints to predict ideology\footnote{Right wing authoritarian users have the largest proportion of comments/posts in r/AgainstHateSubreddits which is contrary to expectation. A brief investigation suggests that this may be because users from r/AgainstHateSubreddits have targeted r/PoliticalCompassMemes due to the posts of the more conservative users. Consequently, some conservative users comment in r/AgainstHateSubreddits to engage in argument.}.  

Of greater concern is the fact that interactions with subreddits that do not pertain to politically charged interests exhibit substantial variation with respect to ideology. For example, as illustrated in the final row, those with authoritarian views seem to have much less interest in movies, video games or sport; perhaps those with authoritarian views are more committed to their political beliefs and consequently these beliefs form a larger part of their identity leaving them less interested in non-political activities. The upshot here is that digital footprints indicating something as simple as your interest in movies or sports are informative with respect to ideology. 

Returning now to textual footprints, in Figure \ref{fig.wordcloud} we see some expected results, right wing economic ideology (which is likely correlated with right wing social views) is correlated with frequent talk of guns, taxes and similar topics. There are also some more subtle results which reflect the psychological basis of ideology. We see that right wing views are correlated with using words like “theft”, “felon” and “defence”, a reflection of the fact that conservatives tend to have a larger fear of threat and loss   \citep{Jost2007}. We also see words like “degenerate” and “whore” which picks up on the higher disgust sensitivity of conservatives   \citep{Inbar2011}. Contrastingly, when it comes to left wing economic views, we see many words related to sexuality indicating a much more open discourse on these topics, perhaps linked with lower disgust sensitivity. Concerningly, words with no obvious political connotations correlate just as much as  explicitly political language. `Specific’ appears to correlate with left wing views as strongly as ‘proletariat’ and ‘cute’ as much as ‘queer’. Words indicating ambiguity (`maybe', `sometimes', `confused') and terms related to music (`album', `song') also appear as top linguistic correlates of left wing economic views. This is a frightening reminder of how digital footprints of an entirely non-political nature can be leveraged to infer ideology.

\section{Conclusion}
\label{Discussion}

We have shown that, publicly available, largely non-political digital footprints can be mapped to ideology with greater than baseline accuracy. This broader claim provides validation of existing results but we have also made several novel contributions: First, we show that digital footprints are able to predict the economic dimension of political ideology to a high degree of accuracy even in a multi-class classification problem that includes a centrist class\footnote{Although including a centrist class creates a noisier problem than just a binary, left/right, ideological classification.} Second, our result suggest that digital footprints are better predictors of economic ideology than social ideology. Third, textual data from a corpus of discussion forum posts is less predictive of political ideology than forum interaction data.



The first two points have important policy upshots. Our results show that digital footprints can predict ideology in a nine-class problem where ideology is a complex variable that contains both an economic and social element. We also show that we can predict economic ideology above baseline accuracy when we account for the existence of centrists, but our predictions are far better when we omit centrists. 


An authoritarian government may want a model that can differentiate the majority of those with conforming ideologies from those with dissenting views. Our results indicate that this is a much more plausible task; models work best when separating two disparate ideologies with the absence of `noisy' ideologies, i.e. those occupying the middle-ground between two extremes. 


Our paper has further illustrated that data from Reddit can be fruitfully utilised in the social sciences. As mentioned, many prior studies in this area relied on Facebook data and linked digital footprints such as Facebook Likes with private traits. The fact that Facebook data is no longer easily accessible to researchers is naturally a poor outcome for this area of inquiry. However, by illustrating Reddit data to be a viable alternative for this sort of inquiry, we hope to revitalise this field. For instance, there are a broad range of Reddit communities where users reveal certain traits (whether through flairs or other means) that can be studied to examine the lins between traits and online behaviour, or traits and behaviour full stop (using online actions as a proxy for general behaviour, i.e. posting in drug related subreddits can be considered a proxy for drug use). 

Further, the data available from Reddit may be useful in other areas of social science research. For instance, using our data set, we could examine the connections between flaired ideology and participation in misogynistic online communities. These kinds of insights into the interactions between all sorts of social behaviours could inspire theories and experiments in a range of disciplines such as psychology, sociology and economics.

\newpage

\bibliographystyle{ecca} 
\bibliography{Bibliography}


\clearpage

\section*{Appendix}
\subsection*{Sections:}
\begin{enumerate}
\item[A.] Flair Recoding
\item[B.] Data in SVD space
\item[C.] Outline of random forests and AdaBoost
\item[D.] Nine-class classification results
\item[E.] Feature Selection
\item[F.] NLP analysis
 
\end{enumerate}

\newpage 

\subsection*{A. Flair Recoding}
\label{Appendix-Recoding}

In the raw user flair data, there are 12 unique flairs recorded: `:CENTG: - Centrist', `:centrist: - Centrist', `:centrist: - Grand Inquisitor', `:left: - Left', `:libright: - LibRight', `:libright2: - LibRight', `:right: - Right', `:libleft: - LibLeft', `:lib: - LibCenter', `:auth: - AuthCenter', `:authleft: - AuthLeft', `:authright: - AuthRight'.

Clearly, some of these are duplicates so we map each of the raw flairs to one of nine flairs like so:

\begin{itemize}
    \item `:CENTG: - Centrist' $\rightarrow$ `centrist'
    \item `:centrist: - Centrist' $\rightarrow$ `centrist'
    \item `:centrist: - Grand Inquisitor' $\rightarrow$ `centrist'
    \item `:left: - Left' $\rightarrow$ `left'
    \item `:libright: - LibRight' $\rightarrow$ `libright'
    \item `:libright2: - LibRight' $\rightarrow$ `libright'
    \item `:right: - Right' $\rightarrow$ `right'
    \item `:libleft: - LibLeft' $\rightarrow$ `libleft'
    \item `:lib: - LibCenter' $\rightarrow$ `libcenter'
    \item `:auth: - AuthCenter' $\rightarrow$ `authcenter'
    \item `:authleft: - AuthLeft' $\rightarrow$ `authleft'
    \item `:authright: - AuthRight' $\rightarrow$ `authright'
\end{itemize}

We use the resulting nine classes as our dependent variable in models for the nine class problem. For the economic problem, we map the nine classes to the one of three classes like so:

\begin{itemize}
    \item `centrist' $\rightarrow$ `center'
    \item `left' $\rightarrow$ `left'
    \item `libright' $\rightarrow$ `right'
    \item `right' $\rightarrow$ `right'
    \item `libleft' $\rightarrow$ `left'
    \item `libcenter' $\rightarrow$ `center'
    \item `authcenter' $\rightarrow$ `center'
    \item `authleft' $\rightarrow$ `left'
    \item `authright' $\rightarrow$ `right'
\end{itemize}

For the social problem:

\begin{itemize}
    \item `centrist' $\rightarrow$ `center'
    \item `left' $\rightarrow$ `center'
    \item `libright' $\rightarrow$ `lib'
    \item `right' $\rightarrow$ `center'
    \item `libleft' $\rightarrow$ `lib'
    \item `libcenter' $\rightarrow$ `lib'
    \item `authcenter' $\rightarrow$ `auth'
    \item `authleft' $\rightarrow$ `auth'
    \item `authright' $\rightarrow$ `auth'
\end{itemize}

\newpage

\subsection*{B. Data in SVD space}
\label{Appendix-svd}

\begin{figure}[!htbp]
\centering
  \fbox{\includegraphics[width=13cm,height=13cm]{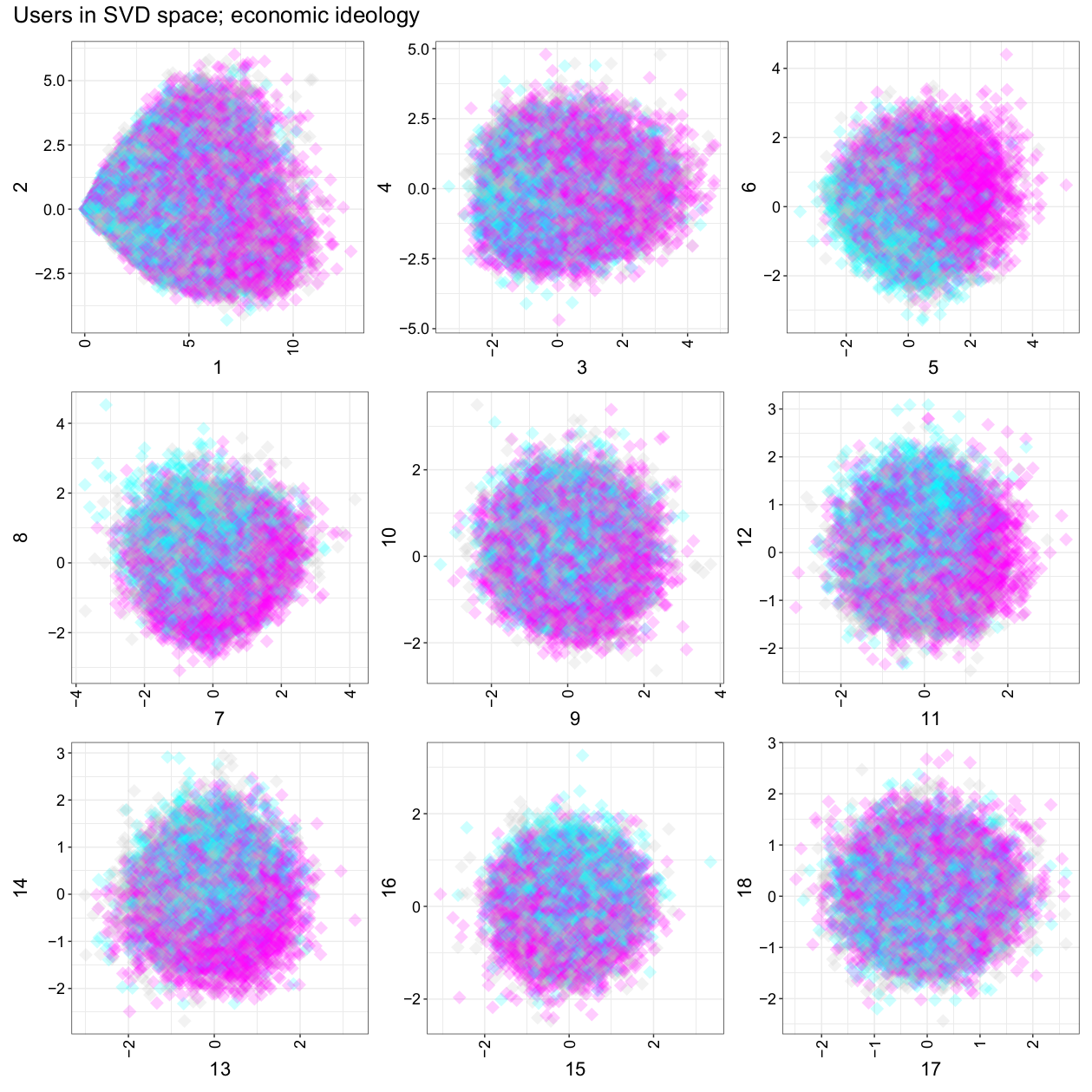}}%
  \caption{ Leftwing (magenta), rightwing (cyan) and centrist (grey) users from the training and validation sets in SVD component space\label{fig.econ_svd}}
\end{figure}

\begin{figure}[!htbp]
\centering
  \fbox{\includegraphics[width=13cm,height=13cm]{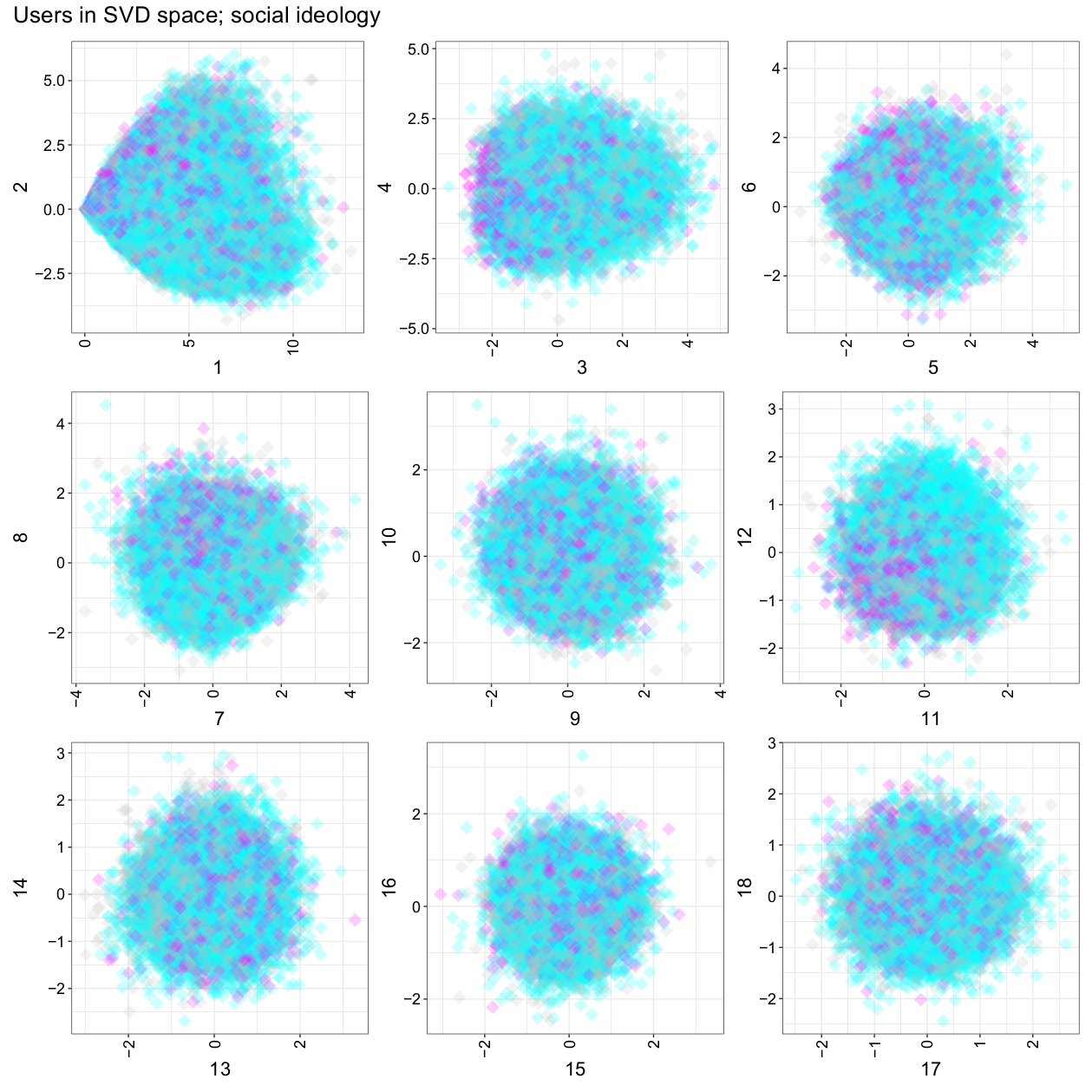}}%
  \caption{Authoritarian (magenta), libertarian (cyan) and centrist (grey)  users from the training and validation sets in SVD component space\label{fig.social_svd}}
\end{figure}

Figures \ref{fig.econ_svd} and \ref{fig.social_svd} show users from the training and validation set in SVD space for SVD components ranging from 1-18 to illustrate the `random' scattering of centrist users. 

\subsection*{C. Outline of random forests and AdaBoost}
\label{Appendix-methodology}

In section \ref{Methodology} we omitted discussion of random forests (with and without an OVR) scheme and AdaBoosting as our none of our strongest models in any modelling problem relied on these supervised learning methods.  Still, we report the results of models that used these algorithms in the supervised learning stage so we have included a brief description of these methods here. 

\subsubsection*{Random forests and OVR random forests}
\label{Appendix-methodology-rf}

A random forest classifier is a a collection of uncorrelated classification trees. We use each of the constituent trees to predict a users ideology and assign user to the ideology which received a plurality of votes from the set of tree models. Each constituent tree may only use a random subsample of the full range of predictors at a given split to ensure that trees remain uncorrelated \citep{10.5555/2517747}. Using an ensemble of tree predictors provides a model with lower variance than an individual classification tree \citep{10.5555/2517747}. Each tree uses a bootstrapped sample of size $n$ as its training set. 

A classification tree recursively partitions the $p$-dimensional feature space into non-overlapping subsets. Binary partitions are made along a single variable to minimise some quantity measuring the impurity (heterogeneous class membership) of the resulting partitions. The splitting process stops when certain criterion are reached. Predictions are made by assigning a new observation to the plurality class of the partition in feature space in which the observation resides. 

The trees in our random forest model make partitions to minimise Gini Impurity \citep{10.5555/2517747}. The Gini Impurity for a given partition, $m$ is given by:

$$ G_m = \sum^K_{k=1} p_{mk}(1-p_{mk})$$

Where $p_{mk}$ refers to the proportion of observations in partition $m$ belonging to class $k$. We experimented with uniform sample weights, $w_i$ and balanced weights where the $w_i$ are assigned to be inversely proportional to class frequency within the relevant bootstrap sample, i.e. $w_i = \frac{n_B}{n_{B, i}}$ where $n_B$ is the number of samples in the relevant bootstrap sample and $n_{B, i}$ is the number of samples of the same class as observation $i$ in $n_B$. Consequently, $p_{mk} = \frac{\sum_{x_i \in m} w_i I(y_i = k)}{\sum_{x_i \in m} w_i}$.

Consider a candidate split of our feature space $X$ into: $P_1 = \{X: x_p < a\}$ and $P_2 = \{X: x_p \ge a\}$. The Gini Impurity of the resulting partitions is given by:

$$n_1 \cdot G_1 + n_2 \cdot G_2 $$
$$= n_1 \cdot \sum^K_{k=1} p_{1k}(1-p_{1k}) + n_2 \cdot \sum^K_{k=1} p_{2k}(1-p_{2k}) $$

Where $n_1$ refers to the number of observations in $P_1$; $|\{X: x_p < a\}|$, and $n_2$ refer to the number of observations in $P_2$; $|\{X: x_p \ge a\}|$. $x_p$ is chosen from the set of predictors and $a$ is chosen from the set of values in the range of $x_p$ to minimise Gini Impurity. 

We specify a minimum bucket and minimum samples stopping criteria. The minimum bucket specification requires that a minimum number of samples be included in each partition. The minimum sample criterion requires that a partition with less samples than the minimum not be split again. 

Random forests are capable of multi-class classification but an OVR estimation scheme as detailed in Section \ref{Methodology-ovr-logreg} can also be applied to random forest models and sometimes outperforms standard random forests \citep{Adnan2015}. Continuing with the economic-ideology classification problem as an example, $y_i \in \{\text{left, center, right}\}$, we would implement an OVR random forest scheme by training three models: a random forest that predicts being left versus not left, center versus not center and right versus not right. 

\subsubsection*{AdaBoost}
\label{Appendix-methodology-adaboost}

AdaBoost is a boosting algorithm that produces a classifier from a set of weak classifiers (in our case, decision trees). A sequence of stumps, $T^{(1)},\dots,T^{(M)}$ (decision trees of depth 1) are created where the errors of $m_{\text{th}}$ tree impact the structure of the $m+1_{\text{th}}$ tree. Initially, all samples are weighted equally. The first model is the decision stump that best partitions the data according to the Gini index (see Section \ref{Appendix-methodology-rf}). The stump's input to the final classification decision is determined by the accuracy with which the stump partitions the training set (impurity metrics account for each samples weight). Finally, the sample weights of observations that were incorrectly classified are increased whilst the weights of those correctly classified are decreased. The magnitude of increases and decreases to sample weights depends on the weight of the most recently trained constituent stump. The next stump is defined by the split that yields the strongest weighted accuracy in its partitioning of the data. This process continues until we have $M$ classifiers working in unison. A technical description of the SAMME multi-class AdaBoost algorithm \citep{Zhu2006} is provided below:

\textbf{Training:}

\begin{enumerate}
    \item Initialise observation weights, $w_i$: $\forall i \in [1,\,n]: w_i := \frac{1}{n}$ 
    \item For $m = 1,\dots, M:$
    \begin{enumerate}
        \item Fit tree stump $T^{(m)}(x_i)$ to training data using Gini index on observations with observations weighted by $w_i$ ((weighted Gini)).
        \item Calculate the error of $T^{(m)}(x_i)$: $\epsilon^{(m)} =  \cfrac{\sum^n_{i=1} w_i I(y_i \not = T^{(m)}(x_i))}{\sum^n_{i=1}w_i}$.
        \item Calculate the weight of $T^{(m)}(x_i)$'s classification in the ensemble classifier where $\eta \le 1$ is the specified learning rate: $\alpha^{(m)} = \eta \cdot \log \cfrac{1-\epsilon^{(m)}}{\epsilon^{(m)}} + \log(K-1)$.
        \item Update the observation weights: $\forall i \in [1,\,n]: w_i := w_i \cdot e^{\alpha^{(m)} I(c_i \not = T^{(m)}(x_i))}$.
        \item Re-normalise observation weights:  $\forall i \in [1,\,n]: w_i := \cfrac{ w_i}{\sum^n_{j=1} w_j}$.
\end{enumerate}
\end{enumerate}   
    
\textbf{Prediction:}

To estimate the class, $C(.)$, of a new observation, $j$, we take the weighted vote of our $M$ tree stumps (votes are weighted by classifier weight $\alpha^{(m)}$) and assign observation $j$ to the class, $k$, which received a plurality of votes:

$$C(x_j) =  \underset{k}{\mathrm{argmax}} \sum^M_{m=1} \alpha^{(m)} I(T^{(m)}(x_j) = k)$$

Rojas provides lighter explanation of the AdaBoost algorithm for binary classification tasks \citep{rojas2009adaboost}.

\subsection*{D. Nine-class classification results} 
\label{Appendix-nineclass}

The results of the Nine-class classification results are reported in Table \ref{table.results_2} illustrates the accuracy and ROC-AUC (where applicable) of all models in the nine-class classification problem.

\begin{table}[!ht]
  \centering
  \caption{Nine-class classification of ideology \\
  \scriptsize  N refers to the total amount of samples divided between training, validation and testing sets. AUC refers to weighted ROC-AUC}
        \scalebox{0.70}{
    \begin{tabular}{llll}
    \makebox[0pt][l]{\textbf{Nine-class classification}} &       &       &  \\
    Model & Accuracy & AUC   & N \\
    \midrule
    \makebox[0pt][l]{\textit{Features: user int}} &       &       &  \\
    \midrule
    ZeroR & 20.81\% &       & 80,961 \\
    Multinomial logit ($\ell_1$) & 29.10\% & 74.20\% & 80,961 \\
    Random forest & 29.23\% & 67.96\% & 80,961 \\
    AdaBoost & 29.98\% & 68.59\% & 80,961 \\
    OVR Random Forest & 30.56\% & 68.51\% & 80,961 \\
    OVR Logit ($\ell_1$) & 34.63\% & 74.18\% & 80,961 \\
    OVR Logit  & 34.69\% & 74.15\% & 80,961 \\
    Multinomial logit & 34.82\% & 74.61\% & 80,961 \\
          &       &       &  \\
    \makebox[0pt][l]{\textit{Features: textual features}} &       &       &  \\
    \midrule
    Linear SVC; w2v & 15.96\% &       & 87,547 \\
    ZeroR & 19.84\% &       & 87,547 \\
    Linear SVC; tf-idf & 24.32\% &       & 87,547 \\
    Linear SVC; combined & 24.40\% &       & 87,547 \\
          &       &       &  \\
    \makebox[0pt][l]{\textit{Features: combined}} &       &       &  \\
    \midrule
    ZeroR & 20.87\% &       & 78,348 \\
    Linear SVC & 33.80\% &       & 78,348 \\
    OVR Logit ($\ell_1$) & 34.59\% & 73.97\% & 78,348 \\
    \end{tabular}}
  \label{table.results_2}
\end{table}%

\subsection*{ E. Feature Selection} 
\label{Appendix-feat-selection}

The relevant features identified for users' economic and social ideology predictions are presented in sections \ref{ecf} and \ref{scf}. The sections contains subreddit features selected for both count and binary user-interaction matrix.

\subsubsection{Economic ideology} \label{ecf}
\textbf{Count features}
\{196, 2meirl4meirl, 4chan, ABoringDystopia, ActualPublicFreakouts, AgainstDegenerateSubs, Anarcho\_Capitalism, Anarchy101, AnimalCrossing, AntiComAction, AntiHateCommunities, AntiLibertarianCringe, Anticommemes, AntifascistsofReddit, AreTheStraightsOK, AsABlackMan, AskLibertarians, AskOuija, AskReddit, BlackPeopleTwitter, BreadTube, CCW, COMPLETEANARCHY, CleanLivingKings, CommunismMemes, ConservativeMemes, Consoom, ContraPoints, CoolAmericaFacts, CoronavirusCirclejerk, CrappyDesign, Cringetopia, Damnthatsinteresting, DankLeft, DeclineIntoCensorship, DemocraticSocialism, DnD, ENLIGHTENEDCENTRISM, EnoughCommieSpam, EnoughMuskSpam, Enough\_Vaush\_Spam, Firearms, FixedPoliticalMemes, ForwardsFromKlandma, FragileWhiteRedditor, Fuckthealtright, Futurology, Gamingcirclejerk, GaySoundsShitposts, GenZAncaps, GenZedong, GenZommunist, GenderCynical, GoGoJoJo, GoldandBlack, Government\_is\_lame, GreenAndPleasant, GunMemes, Hasan\_Piker, JoeRogan, JordanPeterson, JusticeServed, KotakuInAction, LGBTeens, LateStageCapitalism, LateStageImperialism, LeftWithoutEdge, LeftieZ, LeftistGamersUnion, LeopardsAteMyFace, LibertariansBelieveIn, Libright\_Opinion, LivestreamFail, LockdownSkepticism, LouderWithCrowder, LoveForLandlords, MURICA, MadeMeSmile, MarchAgainstNazis, MensLib, Minarchy, MurderedByWords, Music, NFA, NatureIsFuckingLit, NoNewNormal, NoStupidQuestions, Offensivejokes, OurPresident, Overwatch, ParlerWatch, PewdiepieSubmissions, PoliticalCompassMemes, PoliticalHumor, Political\_Revolution, PragerUrine, PresidentialRaceMemes, PublicFreakout, QualitySocialism, RedsKilledTrillions, SRAWeekend, SandersForPresident, SapphoAndHerFriend, SelfAwarewolves, ShitAmericansSay, ShitLiberalsSay, ShitPoliticsSays, Shitstatistssay, Showerthoughts, SmugIdeologyMan, SocialJusticeInAction, Socialism\_101, SocialistRA, SubredditDrama, TIHI, TheLastAirbender, TheLeftCantMeme, TheRightCantMeme, The\_Leftorium, ThisButUnironically, TikTokCringe, ToiletPaperUSA, TopMindsOfReddit, TumblrInAction, TwoXChromosomes, Unexpected, VaushV, VoluntaristMemes, WatchRedditDie, WayOfTheBern, Wellthatsucks, WhitePeopleTwitter, WitchesVsPatriarchy, accidentallycommunist, actuallesbians, alltheleft, antifastonetoss, ar15, askscience, asktransgender, assholedesign, atheism, austrian\_economics, averageredditor, aww, badwomensanatomy, beholdthemasterrace, bi\_irl, bisexual, books, bread\_irl, catsaysmao, centrist, comics, communism101, confidentlyincorrect, coolguides, dankmemes, dataisbeautiful, dndmemes, egg\_irl, ennnnnnnnnnnnbbbbbby, explainlikeimfive, facepalm, fantanoforever, feemagers, forwardsfromgrandma, fragilecommunism, funny, gaming, gatekeeping, gatesopencomeonin, gay\_irl, gifs, goodanimemes, gundeals, gunpolitics, guns, iamverysmart, insaneparents, insanepeoplefacebook, interestingasfuck, leftistvexillology, lgballt, lgbt, libertarianmeme, lostgeneration, marvelstudios, me\_irl, me\_irlgbt, meirl, memes, menwritingwomen, mildlyinfuriating, mildlyinteresting, monarchism, movies, moviescirclejerk, neoliberal, news, nextfuckinglevel, nottheonion, oddlysatisfying, okbuddycapitalist, okbuddyhasan, okbuddyhetero, okbuddyliberty, okbuddyvowsh, onejoke, pics, pointlesslygendered, pokemon, politics, progun, prolife, pussypassdenied, reclassified, religiousfruitcake, science, shitguncontrollerssay, starterpacks, stevenuniverse, stupidpol, tankiejerk, technology, teenagers, todayilearned, traaaaaaannnnnnnnnns, transgendercirclejerk, trees, tucker\_carlson, tumblr, unpopularopinion, videos, walkaway, wallstreetbets, wholesomememes, worldnews, worldpolitics\}

\textbf{Binary features}
\{MakeMeSuffer, LifeProTips, Whatcouldgowrong, AnimalsBeingDerps, bernieblindness, PropagandaPosters, VoluntaristArt, Sigmarxism, trashy, AgainstHateSubreddits, Catholicism, suspiciouslyspecific, AOC, clevercomebacks, darkjokes, seculartalk, MensRights, investing, ScottishPeopleTwitter, gangweed, rightistvexillology, CapHillAutonomousZone, Jreg, comedyheaven, EnoughLibertarianSpam, Jordan\_Peterson\_Memes, MovieDetails, okbuddytankie, Enough\_Sanders\_Spam, indieheads, Classical\_Liberals, Stonetossingjuice, HolUp, YangForPresidentHQ, BikiniBottomTwitter, NintendoSwitch, PlebeianAR, TrumpCriticizesTrump, CapitalismVSocialism, airsoft, AnarchismZ, cyberpunkgame, PleaseCallMeRedScarf, AreTheCisOk, esist, GunPorn, MtF, Art, BisexualTeens, FreeSpeech, AskMen, Drama, longrange, LandlordLove, CombatFootage, europe, Shuffles\_Deck, okbuddyretard, iamatotalpieceofshit, gadgets, woahdude, okbuddydengist, mendrawingwomen, vegan, Qult\_Headquarters, sadcringe, Bad\_Cop\_No\_Donut, fightporn, awfuleverything, blackmagicfuckery, tacticalgear, BrandNewSentence, AbolishTheMonarchy, vexillologycirclejerk, CatholicMemes, DebateAnarchism, pcmasterrace, youtubehaiku, Economics, Bossfight, shitfascistssay, 2020PoliceBrutality, aaaaaaacccccccce, MurderedByAOC, Sino, Trumpgret, hmmm, agedlikemilk, SuddenlyGay, CasualUK, IronFrontUSA, changemyview, AccidentalAlly, AdviceAnimals, Destiny, redditmoment, whiteknighting, bestof, OutOfTheLoop, TooAfraidToAsk, NewDealAmerica, IWW, ShitPoliticalMemes, lewronggeneration, WhereAreAllTheGoodMen, NoFap, Political\_Tumor, SyndiesUnited, thanosdidnothingwrong, TheMonkeysPaw, anarchocommunism, chomsky, BeAmazed, metacanada, television, MilitaryPorn, WatchPeopleDieInside, 2ALiberals, TrollXChromosomes, ProgrammerHumor, sendinthetanks, SocialistGaming, InfowarriorRides, YouShouldKnow, swoletariat, WritingPrompts, QAnonCasualties, CursedGuns, ConservativesOnly, niceguys, Tinder, BasedJustice, quityourbullshit, StarWars, enoughpetersonspam, NonBinary, liberalgunowners, announcements, dontyouknowwhoiam, neoconNWO, LaborwaveAesthetics, AnarchistGenerationZ, ThatsInsane, PoliticalCompass, capitalism\_in\_decay, greentext, FragileMaleRedditor, communism, Minecraft, perfectlycutscreams, Cultural\_Marxism\_irl, okbuddyleftist, MaliciousCompliance, rickandmorty, boomershumor, WinStupidPrizes, menkampf, HumansBeingBros, ak47, MapPorn, CasualConversation, fucktheccp, shitposting, TrueOffMyChest, EnoughTrumpSpam, NoahGetTheBoat, tipofmytongue, Coronavirus, ElizabethWarren, tifu, oldpeoplefacebook, antiwork, IdiotsInCars, Polcompball, modernwarfare, bonehurtingjuice, GatekeepingYuri, StardewValle\} + \{\textbf{Count features}\} - \{GenZedong, dankmemes\}

\subsubsection{Social ideology} \label{scf}
\textbf{Count features}
\{2balkan4you, 4chan, ATBGE, AdviceAnimals, AgainstDegenerateSubs, AgainstHateSubreddits, Anarcho\_Capitalism, Anarchy101, ArchitecturalRevival, AreTheStraightsOK, AskLibertarians, AskReddit, Bad\_Cop\_No\_Donut, BeAmazed, BikiniBottomTwitter, BlackPeopleTwitter, Bombstrap, BrandNewSentence, COMPLETEANARCHY, CatholicMemes, Catholicism, ChoosingBeggars, CleanLivingKings, CommunismMemes, Consoom, CrusaderKings, Damnthatsinteresting, DankLeft, DarkEnlightenment, DebateCommunism, Drama, EarthPorn, EuropeanSocialists, Firearms, For\_Slavs, Futurology, Gamingcirclejerk, GenZedong, GoGoJoJo, GoldandBlack, HistoryMemes, HumansBeingBros, IdiotsInCars, InformedTankie, Jreg, JucheGang, Kaiserposting, Kaiserreich, KidsAreFuckingStupid, LeopardsAteMyFace, LibertariansBelieveIn, LifeProTips, LoveForLandlords, MadeMeSmile, MapPorn, MemriTVmemes, MovieDetails, MurderedByWords, Music, NatureIsFuckingLit, NintendoSwitch, NoStupidQuestions, Offensivejokes, OrthodoxChristianity, PleaseCallMeRedScarf, Polcompball, PoliticalCompass, PoliticalCompassMemes, ProgrammerHumor, PropagandaPosters, PublicFreakout, SapphoAndHerFriend, SelfAwarewolves, ShitLiberalsSay, Shitstatistssay, Showerthoughts, Sino, SyrianCirclejerkWar, TIHI, TNOmod, TheLastAirbender, TheLeftCantMeme, Tinder, ToiletPaperUSA, TradWave, TraditionalCatholics, Unexpected, VoluntaristMemes, WTF, WatchPeopleDieInside, WatchRedditDie, Wellthatsucks, WhitePeopleTwitter, WinStupidPrizes, agedlikemilk, arabfunny, askscience, assholedesign, averageredditor, aww, bi\_irl, bisexual, blackmagicfuckery, centrist, changemyview, comics, communism, communism101, dataisbeautiful, eu4, europe, explainlikeimfive, facepalm, funny, gaming, gatekeeping, gatesopencomeonin, gifs, gunpolitics, heraldry, hoi4, imaginarymaps, insanepeoplefacebook, interestingasfuck, kingsnottrash, lgbt, liberalgunowners, libertarianmeme, libertarianunity, marvelstudios, me\_irl, me\_irlgbt, memes, mildlyinfuriating, mildlyinteresting, monarchism, movies, nba, neoliberal, news, nextfuckinglevel, nfl, nottheonion, oddlysatisfying, okbuddyretard, paleoconservative, paradoxplaza, pcmasterrace, personalfinance, pics, polandball, politics, progun, reclassified, rightistvexillology, russia, science, sendinthetanks, shittymoviedetails, space, starterpacks, stupidpol, technology, television, tifu, todayilearned, traaaaaaannnnnnnnnns, trees, tucker\_carlson, tumblr, unpopularopinion, vexillology, victoria2, videos, virginvschad, wallstreetbets, wholesomememes, worldnews\}

\textbf{Binary features}
\{AskHistorians, FuckYouKaren, AntiHateCommunities, Fuhrerreich, Abolishtherepublic, GenZommunist, austriahungary, CrusadeMemes, Whatcouldgowrong, leftistvexillology, trippinthroughtime, geopolitics, TwoXChromosomes, Chodi, trashy, 8ValuesMemes, suspiciouslyspecific, clevercomebacks, guns, 40kLore, Enough\_Vaush\_Spam, LateStageImperialism, ScottishPeopleTwitter, LibertarianPartyUSA, FellowKids, actuallesbians, 2MiddleEast4you, Apustaja, BanPitBulls, iamverysmart, fragilecommunism, Technocracy, WesternCivilisation, RedsKilledTrillions, ShittyLifeProTips, Classical\_Liberals, Cringetopia, SubredditDrama, FragileWhiteRedditor, rareinsults, ENLIGHTENEDCENTRISM, sabaton, HolUp, BetterEveryLoop, fantanoforever, food, twrmod, SocialJusticeInAction, shittyfoodporn, CapitalismVSocialism, terriblefacebookmemes, USMonarchy, Eyebleach, teenagers, AskEurope, hoi4modding, whatisthisthing, blursedimages, IslamicHistoryMeme, okbuddyhetero, totalwar, UpliftingNews, GetMotivated, Art, The\_Cabal, BeardTube, AbruptChaos, AskMen, fakehistoryporn, TheRightCantMeme, mountandblade, azerbaijan, Shuffles\_Deck, iamatotalpieceofshit, gadgets, buildapc, woahdude, TikTokCringe, AmItheAsshole, technicallythetruth, antifastonetoss, RoughRomanMemes, Socialism\_101, DIY, GenZAncaps, nevertellmetheodds, JusticeServed, DebateAnarchism, toptalent, youtubehaiku, Economics, AlternateHistory, instant\_regret, Rebornyouth, zelda, ParadoxExtra, maybemaybemaybe, CryptoCurrency, weed, LSD, DnD, EDC, antipornography, redditmoment, GunMemes, OutOfTheLoop, Warthunder, islam, pointlesslygendered, TooAfraidToAsk, forwardsfromgrandma, 196, nonononoyes, iamverybadass, meirl, NoFap, thanosdidnothingwrong, GermanWW2photos, Government\_is\_lame, LibertarianLeft, metacanada, Izlam, 2ALiberals, MilitaryPorn, okbuddyliberty, dndmemes, Breath\_of\_the\_Wild, Palestine, ar15, CFB, WritingPrompts, Minarchy, YouShouldKnow, XiIsFinished, RoastMe, DrewDurnil, AteTheOnion, AskBalkans, instantkarma, AskOuija, egg\_irl, catsaysmao, pokemon, badunitedkingdom, CrappyDesign, syriancivilwar, dontyouknowwhoiam, therewasanattempt, ABoringDystopia, LaborwaveAesthetics, ThatsInsane, accidentallycommunist, menwritingwomen, sbubby, photoshopbattles, NFA, Libright\_Opinion, perfectlycutscreams, AntiLibertarianCringe, austrian\_economics, PragerUrine, ModernPropaganda, MaliciousCompliance, rickandmorty, BritishNationalism, SmashBrosUltimate, JoeRogan, specializedtools, Overwatch, Coronavirus, LateStageCapitalism, ColonisingReddit, MonarchoSocialism, kaiserredux, DunderMifflin, sports, prolife, coolguides, bonehurtingjuice, insaneparents\} + \{\textbf{Count features}\} - \{PoliticalCompassMemes\}

\subsection*{D. NLP analysis} 

The transformer models are a new, revolutionary class of deep learning models that use attention mechanisms to understand complex interconnectedness inherent within language structures. We employed the Sentence transformer model \citep{reimers-2019-sentence-bert, song2020mpnet} to retrieve sentence embeddings of the text. Transformer models such as these are trained on more than 1 billion sentence pairs and are fine-tuned to the task of finding sentences with high cosine similarity. Thus, the sentence transformer model is a good fit for the task of clustering similar information.

The user comments dataset contained 87,547 unique users. The user interaction matrix resulting from these users contained 7,479,438 comments from 78,348 users across 68,231 subreddits. Over 15 million sentences were extracted from this dataset, with over 90\% of these containing $<$25 word tokens. The comments corresponding to only the subreddits that were previously selected using feature selection algorithm on the user interaction density matrix were retained. Further, we included both average and max pooling for the sentence embeddings in these subreddits as features per user to train the model. The results are presented in the table \ref{table.results_nlp_fs} below. 

\begin{table}[!htbp] 
  \centering
  \caption{Binary and multi-class classification of economic and social ideology using text\\ 
\scriptsize  N refers to the total amount of samples divided between training, validation and testing sets. In the multi-class problem AUC refers to weighted ROC-AUC, Mean/Max refers to average and max pooling of sentence embeddings}
      \scalebox{0.70}{
    \begin{tabular}{lllllllll}
    \makebox[0pt][l]{\textbf{Multi-class economic classification}} &       &       &       &       & \makebox[0pt][l]{\textbf{Multi-class social classification}} &       &       &  \\
    Model & Accuracy & AUC   & N     &       & Model & Accuracy & AUC   & N \\
\cmidrule{1-4}\cmidrule{6-9}   
    \makebox[0pt][l]{\textit{Features: Text filtered by feature selection on binary data}} &       &       &       &       &       &       &       &  \\
    \\
    \cmidrule{1-4}\cmidrule{6-9} XGBoost (Mean) & 40.91\% & 58.73\% & 77,066 &       & XGBoost (Mean) & 42.48\% &    58.67\%   & 77,208\\
    XGBoost (Max) & 37.72\% & 54.88\% & 77,066  &       & XGBoost (Max) & 38.95\% &    54.53\%   & 77,208 \\
    \\
    \cmidrule{1-4}\cmidrule{6-9}  
    \makebox[0pt][l]{\textit{Features: Text filtered by feature selection on count data}} &       &       &       &       &       &       &       &  \\
    \\
    \cmidrule{1-4}\cmidrule{6-9} XGBoost (Mean) & 41.08\% & 58.94\% & 76071 &       & XGBoost (Mean) & 42.78\% &    59.13\%   & 76,279 \\
    XGBoost (Max) & 36.53\% & 53.99\% & 76071 &       & XGBoost (Max) & 39.5\% &    55.05\%   & 76,279 \\
    
    \end{tabular}}
  \label{table.results_nlp_fs}%
\end{table}%

The transformer model with average pooling of all the embeddings has a similar performance to Glove embeddings for the use case of economic classification. The transformer model has a lower accuracy for social classification when considered at a threshold of 0.5, but has a similar AUC as the Glove embeddings model for this use case. The tf-idf model performs the best as it is most efficient in filtering out the noise and capturing words correlated to ideology. We also compared the performance of the models on textual data based on feature selection performed earlier, i.e the count data based features (subreddits) and the binary data based features (subreddits). The count based features are almost a subset of the binary based features. The models using the text data filtered on these two subsets of subreddit features showed similar performance, suggesting that no new information is added by the additional features introduced for binary data.

We sought to filter out words in the corpus that are strongly aligned with a economic political ideology. A study by Preo$ț$iuc-Pietro et al. \citep{preotiuc-pietro-etal-2017-beyond} presented an approach for the prediction of political preference based on language used in social media, and also provided a mapping of political ideology groups to most frequently used unigrams per group. Leveraging their work, we utilise their lexicons for the ideology groups at the two extreme ends of the spectrum - very conservative and very liberal - and convert these lexicons into right and left leaning Glove word embeddings respectively. Our political ideology axis is defined by the difference between these two vector embeddings. The embeddings generated from user comments is then compared against this ideology axis using cosine similarity, and this comparison enables us to filter out words that are strongly associated with either end of the political ideology spectrum.

\begin{figure}[!htbp]
\centering
  \includegraphics[width=12cm,height=6cm]{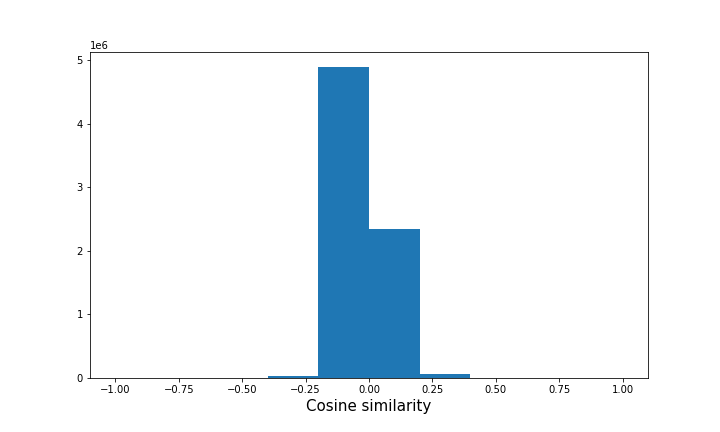}%
  \caption{Cosine similarity of user comments against economic political ideology axis (using Glove embedding) \label{fig.ideology_cosine}}
\end{figure}

The results of this comparison are summarized as a histogram in Figure \ref{fig.ideology_cosine}. The first and last bins of the histogram map to high cosine similarity. Our analysis indicates that a vast majority of words in our corpus do not align strongly with either ends of the economic political ideology spectrum.

\appendix
\section*{Open Source Software References}
\label{Appendix-software}

The analysis underpinning this paper was conducted using multiple R and Python packages. In particular, we used tidyverse  \citesupp{tidyverse}, patchwork  \citesupp{patchwork}, arrow  \citesupp{arrow}, viridis  \citesupp{viridis}, kableExtra  \citesupp{kableExtra}, Matrix  \citesupp{Matrix}, tm  \citep{tm}, wordcloud  \citesupp{wordcloud}, tidytext  \citesupp{tidytext}, gridExtra  \citesupp{gridextra} and ggplotify  \citesupp{ggplotify}  R packages and the pandas  \citesupp{Pandas2}, NumPy  \citesupp{numpy}, SciPy  \citesupp{SciPy}, SciKit-Learn  \citesupp{scikit-learn}, nltk  \citep{nltk}, re  \citesupp{re}, gensim  \citesupp{gensim}, bs4\footnote{\url{https://pypi.org/project/beautifulsoup4/}}, zeugma\footnote{\url{https://zeugma.readthedocs.io/en/latest/}} , pmaw\footnote{\url{https://pypi.org/project/pmaw/}}, PRAW\footnote{\url{https://praw.readthedocs.io/en/stable/}} and 
prawcore\footnote{\url{https://pypi.org/project/prawcore/}} Python packages.

\bibliographystylesupp{ecca} 
\bibliographysupp{Bibliography} 
 \end{document}